# Technical Report: CSVM Ecosystem

## Using CSVM format in various scientific fields.

Frédéric Rodriguez [a,b,*]

[a] CNRS, Laboratoire de Synthèse et Physico-Chimie de Molécules d'Intérêt Biologique, LSPCMIB, UMR-5068, 118 Route de Narbonne, F-31062 Toulouse Cedex 9, France.
[b] Université de Toulouse, UPS, Laboratoire de Synthèse et Physico-Chimie de Molécules d'Intérêt Biologique, LSPCMIB, 118 route de Narbonne, F-31062 Toulouse Cedex 9, France.

## Abstract

The CSVM format is derived from CSV format and allows the storage of tabular like data with a limited but extensible amount of metadata. This approach could help computer scientists because all information needed to uses subsequently the data is included in the CSVM file and is particularly well suited for handling RAW data in a lot of scientific fields and to be used as a canonical format. The use of CSVM has shown that it greatly facilitates: the data management independently of using databases; the data exchange; the integration of RAW data in dataflows or calculation pipes; the search for best practices in RAW data management. The efficiency of this format is closely related to its plasticity: a generic frame is given for all kind of data and the CSVM parsers don't make any interpretation of data types. This task is done by the application layer, so it is possible to use same format and same parser codes for a lot of purposes. In this document some implementation of CSVM format for ten years and in different laboratories are presented. Some programming examples are also shown: a Python toolkit for using the format, manipulating and querying is available. A first specification of this format (CSVM-1) is now defined, as well as some derivatives such as CSVM dictionaries used for data interchange. CSVM is an Open Format and could be used as a support for Open Data and long term conservation of RAW or unpublished data.

* *Corresponding authors.*
CNRS, Laboratoire de Synthèse et Physico-Chimie de Molécules d'Intérêt Biologique, LSPCMIB, UMR-5068, 118 Route de Narbonne, F-31062 Toulouse Cedex 9, France.
Tel.: þ33 (0) 5 61556486; fax: þ33 (0) 5 61556011.
E-mail address: *Frederic.Rodriguez@univ-tlse3.fr* (F. Rodriguez).

# 1. The basics: table, file, annotations and a programming sample

This CSVM sample is relative to the field of medicinal chemistry, *courtesy from* Pascal Hoffmann and coll [1] [2]. This is the case of a collection of 80 molecular files (af01.mdl, af02.mdl, …, af80.mdl) in a directory. Each file stores the molecular 2D coordinates (formula) for a given compound, using a given format (here MDL Molfile format [3]).

*Figure 1. –A view [4] of the molecule's collection.*

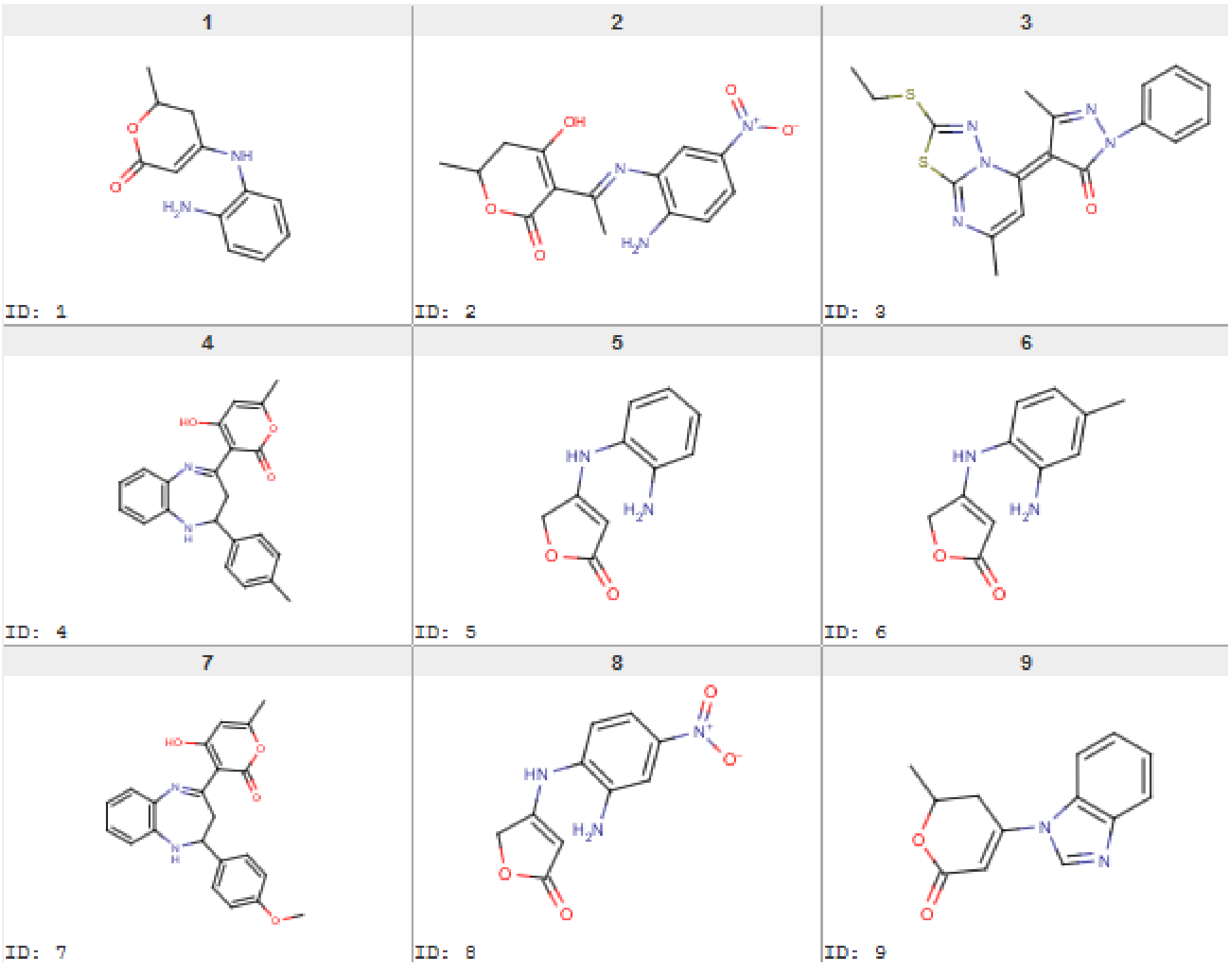

We wanted to store a table of these molecules, including the molecule's number (*ID*), an identifier (*ident*), the chemist related to the compound synthesis (*lab*), an amount (*vrac*) available of the compound, another compound identifier (*rprod*), the related laboratory notebook identifier (*rlab*), the molecular file itself (*molfile*) including a relative and the compound formula (*smi*) written in another format (here SMILES [5] chemical format) and included in a cell of the table:

*Figure 2. –A view of the collection written in a table, only rows [1..10] and [78..80] are shown.*

| *ID* | *ident* | *lab* | *vrac* | *rprod* | *rlab* | *molfile* | *smi* |
|---|---|---|---|---|---|---|---|
| 1 | af01 | hoffmann | 114 | AF01 | MAT1 | mols\af01.mdl | C1C(OC(=O)C=C1Nc1ccccc1N)C |
| 2 | af02 | hoffmann | 85 | AF02 | MAT2 | mols\af02.mdl | c1(c(ccc(c1)[N+](=O)[O-])N)/N=C(/C1=C(O)CC(OC1=O)C)\C |
| 3 | af03 | hoffmann | 60 | AF03 | FM401 | mols\af03.mdl | n12nc(sc1nc(C)c/c/2=C/1\C(=O)N(N=C1C)c1ccccc1)SCC |
| 4 | af04 | hoffmann | 50 | AF04 | FM5b | mols\af04.mdl | c1(ccc(cc1)C)C1Nc2c(N=C(c3c(=O)oc(cc3O)C)C1)cccc2 |
| 5 | af05 | hoffmann | 100 | AF05 | MAT5 | mols\af05.mdl | C1(=CC(=O)OC1)Nc1ccccc1N |
| 6 | af06 | hoffmann | 71 | AF06 | MAT6 | mols\af06.mdl | C1(=CC(=O)OC1)Nc1ccc(cc1N)C |
| 7 | af07 | hoffmann | 60 | AF07 | FM500 | mols\af07.mdl | c1(ccc(cc1)OC)C1Nc2c(N=C(c3c(=O)oc(cc3O)C)C1)cccc2 |
| 8 | af08 | hoffmann | 50 | AF08 | MAT8 | mols\af08.mdl | C1(=CC(=O)OC1)Nc1ccc(cc1N)[N+](=O)[O-] |
| 9 | af09 | hoffmann | 60 | AF09 | MAT38 | mols\af09.mdl | c12c(cccc2)n(cn1)C1=CC(=O)OC(C1)C |
| 10 | af10 | hoffmann | 45 | AF10 | MAT39 | mols\af10.mdl | c12c(ccc(c2)C)n(cn1)C1=CC(=O)OC(C1)C |
| . . . | | | | | | | |

1 L. Hammal, S. Bouzroura, C. André, B. Nedjar-Kolli, P. Hoffmann (2007) Versatile Cyclization of Aminoanilino Lactones: Access to Benzotriazoles and Condensed Benzodiazepin-2-thione. Synthetic Commun., 37:3, 501-511.

2 M .Fodili, M. Amari, B. Nedjar-Kolli, B. Garrigues, C. Lherbet, P. Hoffmann (2009) Synthesis of Imidazoles from Ketimines Using Tosylmethyl Isocyanide (TosMIC) Catalyzed by Bismuth Triflate. Lett. Org. Chem., 6, 354-358.

3 Chemical table file [Wikipedia] - http://en.wikipedia.org/wiki/Chemical_table_file

4 ChemAxon marvinView - http://www.chemaxon.com/products/marvin/marvinview/

5 Chemical file format [Wikipedia] - http://en.wikipedia.org/wiki/Chemical_file_format

| | | | | | | | |
|---|---|---|---|---|---|---|---|
| 78 | af78 | hoffmann | 35 | AF78 | FM406 | mols\af78.mdl | c1(ccccc1)N1C(=O)/C(=c\2/n3c(nc(c2)C)sc2c3ccc(c2)OC)/C(=N1)C |
| 79 | af79 | hoffmann | 43 | AF79 | FM267 | mols\af79.mdl | c1(ccccc1)n1c(c(c(n1)C)C(=O)/C=C(\Nc1sc2c(n1)c(ccc2)C)/C)O |
| 80 | af80 | hoffmann | 45 | AF80 | FM257 | mols\af80.mdl | s1cccc1Cn1c(c(nc1)C)c1c(O)cc(C)oc1=O |

Then we wanted to write the table: *i)* as a CSV derived file for data and *ii)* with a block for metadata:

*Figure 3. –The corresponding CSVM file (rows [1..10] and [78..80]).*

```
01	af01	hoffmann	114	AF01	MAT1	mols\af01.mdl	C1C(OC(=O)C=C1Nc1ccccc1N)C
02	af02	hoffmann	85	AF02	MAT2	mols\af02.mdl	c1(c(ccc(c1)[N+](=O)[O-])N)/N=C(/C1=C(O)CC(OC1=O)C)\C
03	af03	hoffmann	60	AF03	FM401	mols\af03.mdl	n12nc(sc1nc(C)c/c/2=C/1\C(=O)N(N=C1C)c1ccccc1)SCC
04	af04	hoffmann	50	AF04	FM5b	mols\af04.mdl	c1(ccc(cc1)C)C1Nc2c(N=C(c3c(=O)oc(cc3O)C)C1)cccc2
05	af05	hoffmann	100	AF05	MAT5	mols\af05.mdl	C1(=CC(=O)OC1)Nc1ccccc1N
06	af06	hoffmann	71	AF06	MAT6	mols\af06.mdl	C1(=CC(=O)OC1)Nc1ccc(cc1N)C
07	af07	hoffmann	60	AF07	FM500	mols\af07.mdl	c1(ccc(cc1)OC)C1Nc2c(N=C(c3c(=O)oc(cc3O)C)C1)cccc2
08	af08	hoffmann	50	AF08	MAT8	mols\af08.mdl	C1(=CC(=O)OC1)Nc1ccc(cc1N)[N+](=O)[O-]
09	af09	hoffmann	60	AF09	MAT38	mols\af09.mdl	c12c(cccc2)n(cn1)C1=CC(=O)OC(C1)C
10	af10	hoffmann	45	AF10	MAT39	mols\af10.mdl	c12c(ccc(c2)C)n(cn1)C1=CC(=O)OC(C1)C
...
78	af78	hoffmann	35	AF78	FM406	mols\af78.mdl	c1(ccccc1)N1C(=O)/C(=c\2/n3c(nc(c2)C)sc2c3ccc(c2)OC)/C(=N1)C
79	af79	hoffmann	43	AF79	FM267	mols\af79.mdl	c1(ccccc1)n1c(c(c(n1)C)C(=O)/C=C(\Nc1sc2c(n1)c(ccc2)C)/C)O
80	af80	hoffmann	45	AF80	FM257	mols\af80.mdl	s1cccc1Cn1c(c(nc1)C)c1c(O)cc(C)oc1=O

#TITLE	
#HEADER	ID	ident	lab	vrac	rprod	rlab	molfile	smi
#TYPE	TEXT	TEXT	TEXT	TEXT	TEXT	TEXT	TEXT	TEXT
#WIDTH	10	10	10	10	10	10	10	10
#META	
```

This is what we call a CSVM file (CSV with Metadata). In this case the field separator is a TAB (red arrows) but the CSVM specification allows any character. The data block is the same than shown in Figure 2 but it is followed by rows beginning with # character and defining a metadata block.

Using flat CSVM files (CSVM-1 specification) the # rows are considered as metadata rows if the # is immediately followed by a known keyword (TITLE, HEADER, TYPE, WIDTH, META). If the keyword is not recognized by the CSVM parser, the row is not processed.

This is also possible to use this property in order to extinct some data rows in the CSVM without deleting them; and to annotate the file adding rows for remarks. The following CSVM illustrates these two cases outside the metadata block (rows 5, 6, 7). The blank lines are not taken in account by the CSVM parser, so they could be used to make the table more readable for humans. As a result, in Figure 4 the compound with ID=3 will not be included in data structure stored in computer's memory.

*Figure 4. –A tagged (remarks) CSVM file.*

```
1
2	01	af01	hoffmann	114	AF01	MAT1	mols\af01.mdl	C1C(OC(=O)C=C1Nc1ccccc1N)C
3	02	af02	hoffmann	85	AF02	MAT2	mols\af02.mdl	c1(c(ccc(c1)[N+](=O)[O-])N)/N=C(/C1=C(O)CC(OC1=O)C)\C
4
5	#=========================================
6	# this record is not verified to date
7	#03	af03	hoffmann	60	AF03	FM401	mols\af03.mdl	n12nc(sc1nc(C)c/c/2=C/1\C(=O)N(N=C1C)c1ccccc1)SCC
8
9	04	af04	hoffmann	50	AF04	FM5b	mols\af04.mdl	c1(ccc(cc1)C)C1Nc2c(N=C(c3c(=O)oc(cc3O)C)C1)cccc2
10	05	af05	hoffmann	100	AF05	MAT5	mols\af05.mdl	C1(=CC(=O)OC1)Nc1ccccc1N
11	06	af06	hoffmann	71	AF06	MAT6	mols\af06.mdl	C1(=CC(=O)OC1)Nc1ccc(cc1N)C
12	07	af07	hoffmann	60	AF07	FM500	mols\af07.mdl	c1(ccc(cc1)OC)C1Nc2c(N=C(c3c(=O)oc(cc3O)C)C1)cccc2
13	08	af08	hoffmann	50	AF08	MAT8	mols\af08.mdl	C1(=CC(=O)OC1)Nc1ccc(cc1N)[N+](=O)[O-]
14	09	af09	hoffmann	60	AF09	MAT38	mols\af09.mdl	c12c(cccc2)n(cn1)C1=CC(=O)OC(C1)C
15	10	af10	hoffmann	45	AF10	MAT39	mols\af10.mdl	c12c(ccc(c2)C)n(cn1)C1=CC(=O)OC(C1)C
16	...
17	78	af78	hoffmann	35	AF78	FM406	mols\af78.mdl	c1(ccccc1)N1C(=O)/C(=c\2/n3c(nc(c2)C)sc2c3ccc(c2)OC)/C(=N1)C
18	79	af79	hoffmann	43	AF79	FM267	mols\af79.mdl	c1(ccccc1)n1c(c(c(n1)C)C(=O)/C=C(\Nc1sc2c(n1)c(ccc2)C)/C)O
19	80	af80	hoffmann	45	AF80	FM257	mols\af80.mdl	s1cccc1Cn1c(c(nc1)C)c1c(O)cc(C)oc1=O
20
21	#TITLE	
22	#HEADER	ID	ident	lab	vrac	rprod	rlab	molfile	smi
23	#TYPE	TEXT	TEXT	TEXT	TEXT	TEXT	TEXT	TEXT	TEXT
24	#WIDTH	10	10	10	10	10	10	10	10
25	#META	
26
```

The metadata block is used to embed the minimal canonical information about the table:

- The `#TITLE` row is for information (one ASCII line) about the table.
- The `#HEADER` row stores for column titles.
- The `#TYPE` row stores for the data types for columns. CSVM lets you define your data types, only fuzzy and very minimal data types are provided: TEXT (textual data), NUMERIC (columns of integer or float values), BOOLEAN (0/1, Y/N, Yes/No you can define your own code).
- The `#WIDTH` row is used to store a value significant of column's width.
- The `#META` row is used to store other/not defined information.

The CSVM parser don’t use the information stored in metadata block: all the information is read as strings and returned as a data structure depending on the programming language used.
Metadata is not interpreted at the parser level and it is a choice. Using this approach makes it possible to add all data types for `#TYPE` keywords, this is the reason of the reduced number of predefined types in CSVM specification. The data type recognition is done only when the data structure is available in memory: this step is uncoupled from the parser step. A side result is that the same parser can be used for all data types.
In the case of Python language, a Class (a csvm_ptr object) is returned with some methods (i.e. csvm_ptr_dump for display the data structure on standard output). The following code (Code 1) shows the contents of the corresponding data structure. These lines are extracted from a Python toolkit (Pybuild) that support CSVM files.

*Code 1. – Beginning of Python CSVM Class code.*

```
class csvm_ptr:
   """
   Follows CSVM specs (v:1.x) for contents of data structure. Standard column
   types are NUMERIC,TEXT,DATE,BOOLEAN. Some of us, use also INTEGER, FLOAT
   for numeric types. Some of us, use also NODE, LINK, IMAGE for web data
   embedded in CSVM files. WIDTHs (10,50 if not set) are for Javascript tables
   and can be omitted.
   *** 1.01/080304/fred
   """
   def __init__(self):
       self.SOURCE = ""        # path/file name of readed CSVM file
       self.CSV = ""           # CSVM or CSV depending of file contents
       self.TITLE_N = 0        #Titles of CSVM file (let for future, only one string used today)
       self.TITLE = ""         # Title of CSVM file
       self.HEADER_N = 0       # Number of data columns titles
       self.HEADER = []        # List of data column titles
       self.TYPE_N = 0         # Number of data columns types (= self.HEADER_N)
       self.TYPE = []          # List of data column types
       self.WIDTH_N = 0        # Number of data columns widths (= self.HEADER_N)
       self.WIDTH = []         # List of data column widths
       self.DATA_R = 0         # Number of data rows
       self.DATA_C = 0         # Number of data columns (= self.HEADER_N)
       self.DATA = []          # String matrix containing data
       self.META = ""          # Meta string
```

The following Python code (Code 2) shows how *1)* to load a CSVM file (using the TAB `"\t"` as delimiter) with the function csvm_ptr_read_extended_csvm, *2)* to display the Class and *3)* to free the memory.

*Code 2. – Loading and printing a CSVM file using Pybuild toolkit.*

```
print "=> A new blank CSVM object"
c = csvm_ptr()
print "=> Populates it with a CSVM file ... "
c = csvm_ptr_read_extended_csvm(c, file_cleanpath("test/hoffmann.csvm"), "\t")
print "=> data dump ... "
c.csvm_ptr_dump(0,0)
print "=> Clear CSVM object"
c.csvm_ptr_clear()
print
```

The corresponding output is shown in Figure 5, the result of csvm_ptr_dump method shows the metadata block (pink text, the order of metadata rows is #WDITH, #TEXT, #HEADER) followed by data rows (all cells are surrounded by brackets). Please notice that the previous annotated CSVM file is used: the row with [03] value in first column {number} is missing.

*Figure 5. – Console output corresponding to Code 1 example, only 11 first data rows and 5 last data rows are displayed.*

```
=> A new blank CSVM object
=> Populates it with a CSVM file ...
=> data dump ...

DUMP: CSVM info {
SOURCE    test\hoffmann.csvm
CSV  CSVM
META      []
TITLE_N   1
TITLE
HEADER_N      15
TYPE_N    15
WIDTH_N   15
0     10    TEXT {number}
1     10    TEXT {name}
2     10    TEXT {plate}
3     10    TEXT {chemist}
4     10    TEXT {amount}
5     10    TEXT {ref_product}
6     10    TEXT {ref_labbook}
7     10    TEXT {id_lab}
8     10    TEXT {id_team}
9     10    TEXT {id_box}
10    10    TEXT {rights}
11    10    TEXT {chr_row_box}
12    10    TEXT {num_col_box}
13    10    TEXT {OpenBabel Symmetry Classes}
14    10    TEXT {smi}
DATA_R    79
DATA_C    15
      79    15
0     [01][af01][cob.1][hoffmann][114][AF01][MAT1][CCC][03][01][L][A][02][6 12 7 13 4 10 15 16 9 5 3 14 8 11 2 1][C1C(OC(=O)C=C1Nc1ccccc1N)C]
1     [02][af02][cob.1][hoffmann][85][AF02][MAT2][CCC][03][01][L][B][02][19 20 10 12 21 13 16 15 18 11 2 17 3 4 8 14 9 5 1 22 6 7][c1(c(ccc(c1)[N+](=O)[O-])N)/N=C(/C1=C(O)CC(OC1=O)C)\C]
2     [04][af04][cob.1][hoffmann][50][AF04][FM5b][CCC][03][01][L][D][02][19 8 7 16 7 8 22 25 15 17 12 24 21 20 14 10 23 18 13 11 5 6 9 4 3 2 1][c1(ccc(cc1)C)C1Nc2c(N=C(c3c(=O)oc(cc3O)C)C1)cccc2]
3     [05][af05][cob.1][hoffmann][100][AF05][MAT5][CCC][03][01][L][E][02][12 5 6 11 7 3 9 13 14 8 4 2 10 1][C1(=CC(=O)OC1)Nc1ccccc1N]
4     [06][af06][cob.1][hoffmann][71][AF06][MAT6][CCC][03][01][L][F][02][12 4 5 11 6 7 9 14 15 8 13 3 10 2 1][C1(=CC(=O)OC1)Nc1ccc(cc1N)C]
5     [07][af07][cob.1][hoffmann][60][AF07][FM500][CCC][03][01][L][G][02][19 9 8 17 8 9 26 25 16 18 15 23 22 21 14 11 24 20 13 12 5 6 10 4 3 2 7 1][c1(ccc(cc1)OC)C1Nc2c(N=C(c3c(=O)oc(cc3O)C)C1)cccc2]
6     [08][af08][cob.1][hoffmann][50][AF08][MAT8][CCC][03][01][L][H][02][13 5 6 12 7 10 8 14 15 11 16 2 9 1 17 3 4][C1(=CC(=O)OC1)Nc1ccc(cc1N)[N+](=O)[O-]]
7     [09][af09][cob.1][hoffmann][60][AF09][MAT38][CCC][03][01][L][A][03][16 15 7 3 11 6 4 8 17 9 12 5 13 14 10 2 1][c12c(cccc2)n(cn1)C1=CC(=O)OC(C1)C]
8     [10][af10][cob.1][hoffmann][45][AF10][MAT39][CCC][03][01][L][B][03][17 16 6 5 11 7 14 8 18 9 12 4 13 15 10 2 1 3][c12c(ccc(c2)C)n(cn1)C1=CC(=O)OC(C1)C]
9     [11][af11][cob.1][hoffmann][43][AF11][MAT36][CCC][03][01][L][C][03][17 16 12 15 9 10 11 8 18 3 7 5 14 13 6 4 1 2][c12c(cc(cc2)Cl)n(cn1)C1=CC(=O)OC(C1)C]
10    [12][af12][cob.1][hoffmann][45][AF12][FM5d][CCC][03][01][L][D][03][26 24 13 6 8 10 27 21 17 18 16 20 19 25 14 12 22 23 15 9 7 5 11 4 3 1 2][c1(c(cccc1)O)C1Nc2c(N=C(c3c(=O)oc(cc3O)C)C1)cccc2]
...
74    [76][af76][cob.1][hoffmann][50][AF76][FM414][CCC][03][01][L][D][11][22 9 14 11 19 13 10 1 20 12 21 18 15 8 17 7 6 16 7 5 6 4 3 2][n12nc(sc1nc(C)c/c/2=C/1\C(=O)N(N=C1C)c1ccccc1)S]
75    [77][af77][cob.1][hoffmann][45][AF77][FM410][CCC][03][01][L][E][11][18 8 6 5 6 8 24 20 17 9 19 26 21 11 15 13 22 23 25 12 7 16 10 14 4 2 3 1][c1(ccccc1)N1C(=O)/C(=c\2/n3c(nc(c2)C)sc2c3ccc(c2)C)/C(=N1)C]
76    [78][af78][cob.1][hoffmann][35][AF78][FM406][CCC][03][01][L][F][11][19 9 6 5 6 9 25 21 18 10 20 27 22 12 16 14 23 24 26 13 8 17 11 15 4 2 3 7 1][c1(ccccc1)N1C(=O)/C(=c\2/n3c(nc(c2)C)sc2c3ccc(c2)OC)/C(=N1)C]
77    [79][af79][cob.1][hoffmann][43][AF79][FM267][CCC][03][01][L][G][11][21 16 7 6 7 16 26 22 27 11 20 9 17 10 19 23 25 14 18 15 24 12 8 13 2 5 1 4 3][c1(ccccc1)n1c(c(c(n1)C)C(=O)/C=C(\Nc1sc2c(n1)c(ccc2)C)/C)O]
78    [80][af80][cob.1][hoffmann][45][AF80][FM257][CCC][03][01][L][H][11][8 6 5 7 15 16 3 12 14 1 19 11 2 21 13 17 18 4 9 20 10][s1cccc1Cn1c(c(nc1)C)c1c(O)cc(C)oc1=O]
}
 done

=> Clear CSVM object
```

# 2. Advanced: combining with databases

CSVM is obviously not a substitute for databases, but it is possible to use a CSVM table for storing simple database schema and database tables. The following sample shows this kind of approach in the case of a chemical inventory. These files were generated after analysis (SQL requests) of a relational database system (Hughes Technologies mQSL) and prior to be converted and injected in another RDBMS (MySQL or PostgreSQL).

## 2.1. Database tables and schema - Chemical inventory

A simple chemical database of two tables and a relation between the column of indice=3 (from 0 to n-1) in table 'assoc' to column of indice=0 in table 'user'. Note that the schema contains heterogeneous data in columns (data from keyword TABLES, or FOREIGN) so all columns titles are set to the same word 'DATA', *courtesy from* Anne Laure Leomant [6] and Nathalie Gouardères.

*Figure 6. – Example of database schema encoded in a CSVM file.*

```
DB      192.168.10.17   anonymous          -       Chemb
TABLE   user            chemb\user.csvm    0       -
TABLE   assoc           chemb\assoc.csvm   0       -
FOREIGN assoc           3                  user    0

#TITLE    Database import schema
#HEADER   KEYWORD    DATA     DATA     DATA     DATA
#TYPE     TEXT       TEXT     TEXT     TEXT     TEXT
#WIDTH    10         10       10       10       10
#META     april/25/2008 15:42:45
```

The database tables are a particular case of CSVM file. Database systems have specific data types (i.e. REAL, INTEGER, SMALLINT …) depending on the records used in tables that are more precise than the generic CSVM type (NUMERIC in this case). But the CSVM approach allows the use of database types in metadata #TYPE row. Also in row #WIDTH it is possible to store the dimension of database fields when it is needed by database type. The following example shows the last 3 lines of CSVM file assoc.csvm corresponding to the `'assoc'` table included in of the previous schema:

*Figure 7. – Example of a database table encoded in a CSVM file.*

```
...
7374   124-63-0   acros    88   12564     CH3ClO2S   1   14-Apr-2006   16:02:07   methane sulfonyl chloride   0.00   100ml
7375   7758-99-8  billault 90   40-3588   CuO4S      1   28-Mar-2006   17:15:13   cuivre sulfate              0.00   500g
7376   142-96-1   acros    48   14969     C8H18O     1   28-Mar-2006   17:20:00   di n-butyl ether            0.00   11

#TITLE   assoc
#HEADER  clef   cas     four   uid   ref    formule   quantite   date   heure   nom    prix   cond
#TYPE    INT    NCHAR   CHAR   INT   CHAR   CHAR      INT        CHAR   TIME    CHAR   CHAR   CHAR
#WIDTH   5      16      16     3     16     18        2          11     8       48     7      16
#META    april/23/2008 10:30:45
```

Notice that other information on databases (i.e. data types for descriptions and conversions) could be also encoded in CSVM files.

*So CSVM, even if it is primarily conceived for handling RAW data is also usable to get a 'flat' view of simple database systems. It could be used as a canonical format to do a lot of transformation between databases, without any modification of CSVM parsers.*

[6] A.L. Leomant. Evaluation d'un SGBDR pour une chimiothèque. Rapport L3 SID, Université Paul Sabatier Toulouse III (2008).

# 3. Indexes, collection of files, documents …

Initially CSVM was used primarily for indexing purposes: file's catalog (filenames/paths and some metadata about the files) rather than as a substitute for spreadsheets tables. The following figure shows this kind of use: the result for a recursive directory list.

*Figure 8. – Example of using a CSVM table for a file catalog (this output is a dump of a CSVM Python class).*

```
DUMP: CSVM info {
SOURCE
CSV CSVM
META    []
TITLE_N 1
TITLE
HEADER_N    3
TYPE_N  3
WIDTH_N 3
0   50  TEXT    {DIR}
1   50  TEXT    {FILE}
2   50  TEXT    {-}
DATA_R  118
DATA_C  3
    118 3
0   [c:\Python26\Lib\site-packages\build][args.py][18:jun:2009]
1   [c:\Python26\Lib\site-packages\build][date.py][19:oct:2006]
2   [c:\Python26\Lib\site-packages\build][dir.py][06:may:2010]
...
115 [c:\Python26\Lib\site-packages\build\_depot\numerix\_dist][__init__.py][16:apr:2008]
116 [c:\Python26\Lib\site-packages\build\_struct][strpdb.py][26:mar:2007]
117 [c:\Python26\Lib\site-packages\build\_struct][__init__.py][31:oct:2006]
}
 done
```

Notice that the last column is filled by the date of creation of files, but the column name is pending (the #HEADER value is a blank) before saving data in a CSVM file. *This is a very interesting feature of CSVM approach related to RAW data: in some circumstances the name of a column is not yet defined but the data must be stored anyway.*

## 3.1. Software pipes - Molecular calculations

CSVM can be useful for launching calculations, you can store in the CSVM the operations to do, including launching external programs, store intermediate results (or index filenames) and complete or add columns from step to steps. Eventually a software component using a CSVM parser can generate from these tables, information (files, parameters) needed by job schedulers/batch processors.
Very often calculations are applied to data files, so this is a particular case of indexing. The interest of CSVM space is that all steps (defining files, parameters, results) could be made easily and in the same data format (ready for use by spreadsheets or scientific data analysis and graphic software).
In the next section, a complete example will be shown (case of Enzyme kinetics). Here a simpler sample issued from [structural biology/rational drug design] science is given. The data shown is a CSVM file, used by a software component added to UCSF Chimera [7] [8] software and used to launch X-Score [9] [10] calculations inside the interface, *courtesy from* Mansi Trivedi [11].
Each calculation uses a ligand and a protein. For a given ligand, one or two torsion angles were selected using the Chimera GUI and different files (`PARAM` column) corresponding to different angular values

7 UCSF Chimera - http://www.cgl.ucsf.edu/chimera/
8 UCSF Chimera--a visualization system for exploratory research and analysis. E.F. Pettersen, T.D. Goddard, C.C. Huang, G.S. Couch, D.M. Greenblatt, E.C. Meng, T.E. Ferrin. *J Comput. Chem.* (2004) 25:13, 1605-1612.
9 Renxiao Wang – http://sw16.im.med.umich.edu/software/xtool/
10 R. Wang, Y. Lu, S. Wang. Comparative Evaluation of 11 Scoring Functions for Molecular Docking. *J. Med. Chem.* (2003) 46:12, 2287-2303.
11 M. Trivedi – *In silico* approach to study protein-ligand binding affinity. Rapport M2P Bioinformatique, Université Paul Sabatier (2009).

('ANG1' and 'ANG2' columns) were generated. For calculation the corresponding log files were added ('LOG' column) in the CSVM table.

*Figure 9. – Calculations driven by a CSVM table.*

```
0    5    4    -    C:/Local/Chimera/calc/ligand_5_4.mol2      C:/Local/Chimera/calc/ligand_5_4_xscore.log      -
1    5    24   -    C:/Local/Chimera/calc/ligand_5_24.mol2     C:/Local/Chimera/calc/ligand_5_24_xscore.log     -
2    5    44   -    C:/Local/Chimera/calc/ligand_5_44.mol2     C:/Local/Chimera/calc/ligand_5_44_xscore.log     -
3    5    64   -    C:/Local/Chimera/calc/ligand_5_64.mol2     C:/Local/Chimera/calc/ligand_5_64_xscore.log     -
4    40   4    -    C:/Local/Chimera/calc/ligand_40_4.mol2     C:/Local/Chimera/calc/ligand_40_4_xscore.log     -
5    40   24   -    C:/Local/Chimera/calc/ligand_40_24.mol2    C:/Local/Chimera/calc/ligand_40_24_xscore.log    -
6    40   44   -    C:/Local/Chimera/calc/ligand_40_44.mol2    C:/Local/Chimera/calc/ligand_40_44_xscore.log    -
7    40   64   -    C:/Local/Chimera/calc/ligand_40_64.mol2    C:/Local/Chimera/calc/ligand_40_64_xscore.log    -
8    75   4    -    C:/Local/Chimera/calc/ligand_75_4.mol2     C:/Local/Chimera/calc/ligand_75_4_xscore.log     -
9    75   24   -    C:/Local/Chimera/calc/ligand_75_24.mol2    C:/Local/Chimera/calc/ligand_75_24_xscore.log    -
10   75   44   -    C:/Local/Chimera/calc/ligand_75_44.mol2    C:/Local/Chimera/calc/ligand_75_44_xscore.log    -
11   75   64   -    C:/Local/Chimera/calc/ligand_75_64.mol2    C:/Local/Chimera/calc/ligand_75_64_xscore.log    -
12   110  4    -    C:/Local/Chimera/calc/ligand_110_4.mol2    C:/Local/Chimera/calc/ligand_110_4_xscore.log    -
13   110  24   -    C:/Local/Chimera/calc/ligand_110_24.mol2   C:/Local/Chimera/calc/ligand_110_24_xscore.log   -
14   110  44   -    C:/Local/Chimera/calc/ligand_110_44.mol2   C:/Local/Chimera/calc/ligand_110_44_xscore.log   -
15   110  64   -    C:/Local/Chimera/calc/ligand_110_64.mol2   C:/Local/Chimera/calc/ligand_110_64_xscore.log   -
16   145  4    -    C:/Local/Chimera/calc/ligand_145_4.mol2    C:/Local/Chimera/calc/ligand_145_4_xscore.log    -
17   145  24   -    C:/Local/Chimera/calc/ligand_145_24.mol2   C:/Local/Chimera/calc/ligand_145_24_xscore.log   -
18   145  44   -    C:/Local/Chimera/calc/ligand_145_44.mol2   C:/Local/Chimera/calc/ligand_145_44_xscore.log   -
19   145  64   -    C:/Local/Chimera/calc/ligand_145_64.mol2   C:/Local/Chimera/calc/ligand_145_64_xscore.log   -

#TITLE     XSCORE For 2 TORSION
#HEADER    IND       ANGL1     ANGL2     PARAM     PDB       LIG       LOG       COF
#TYPE      STRING    STRING    STRING    STRING    STRING    STRING    STRING    STRING
#WIDTH     300       300       300       300       300       300       300       300
```

After calculation, the log files generated by X-SCORE were parsed, converted in CSVM tables, and the tables were indexed (extraction of score values). Then, the score values were aggregated with the upper table (addition of columns). All the process is driven by the table of Figure 9 that contains all information needed for calculation and collection of results. In this case the first column ('IND') is used for store indices of rows but it can be used in *key/values* scheme, as it was done in the following sample:

*Figure 10. – Calculation and parameters driven by a CSVM table, columns PHI and PSI are used for angle values.*

```
1
2   PDB	-	-	dock_test2.pdb	-	-
3   FIXPDB	-	-	e:\test\xscoretest\data	e:\test\xscoretest\input	-
4   FIXMOL2	-	-	e:\test\xscoretest\data	e:\test\xscoretest\input	-
5   SCOREPATHS	-	-	e:\test\xscoretest\input	e:\test\xscoretest\output	-
6   XSCOREIN	-	-	-	-	-
7   ANG	0	-	mol507_0_xscore.mol2	mol507_0_xscore.log	-
8   ANG	20	-	mol507_20_xscore.mol2	mol507_20_xscore.log	-
9   ANG	40	-	mol507_40_xscore.mol2	mol507_40_xscore.log	-
10  ANG	60	-	mol507_60_xscore.mol2	mol507_60_xscore.log	-
11  ANG	80	-	mol507_80_xscore.mol2	mol507_80_xscore.log	-
12  ANG	100	-	mol507_100_xscore.mol2	mol507_100_xscore.log	-
13  ANG	120	-	mol507_120_xscore.mol2	mol507_120_xscore.log	-
14  ANG	140	-	mol507_140_xscore.mol2	mol507_140_xscore.log	-
15  ANG	160	-	mol507_160_xscore.mol2	mol507_160_xscore.log	-
16  ANG	180	-	mol507_180_xscore.mol2	mol507_180_xscore.log	-
17  ANG	200	-	mol507_200_xscore.mol2	mol507_200_xscore.log	-
18  ANG	220	-	mol507_220_xscore.mol2	mol507_220_xscore.log	-
19  ANG	240	-	mol507_240_xscore.mol2	mol507_240_xscore.log	-
20  ANG	260	-	mol507_260_xscore.mol2	mol507_260_xscore.log	-
21  ANG	280	-	mol507_280_xscore.mol2	mol507_280_xscore.log	-
22  ANG	300	-	mol507_300_xscore.mol2	mol507_300_xscore.log	-
23  ANG	320	-	mol507_320_xscore.mol2	mol507_320_xscore.log	-
24  ANG	340	-	mol507_340_xscore.mol2	mol507_340_xscore.log	-
25  ANG	-	-	mol507_xscore.mol2	mol507_xscore.log	-
26
27  #TITLE	X-Score index file
28  #HEADER	KW	PHI	PSI	MOL2	LOG	SCORE
29  #TYPE	STRING	NUMERIC	NUMERIC	STRING	STRING	NUMERIC
30  #WIDTH	50	50	50	50	50	50
31  #META	Test file not definitive format
32
```

The first column labeled 'KW' stores keyword that drives contents of next columns, and the last column ('SCORE') is present and ready to store score values. The CSVM file is used to store parameters/results of calculations (rows 7-25) as well as configuration options of X-SCORE (rows 2-6). This case shows that in the same format space we can store configurations, parameters, results …

# 4. From indexes to collections: files, information and dataflows

The concept was then used for indexes and for tabular data in the same or different collections. For a same catalog, the two next samples show differences (files *vs.* tables) on indexed contents: *i)* the case of a publishing chain (for standard files) and *ii)* the case of a collection of sparse environmental data (for CSVM files). These two examples are issued from scientific works in the field of environmental sciences, but making collections of RAW/processed objects is a general problem in all sciences.

## 4.1. A publishing chain

The process starts from a set of documents (books, articles, reports, various documents) about a particular place (Malause) known as a sedimentation zone in a river (Garonne, France). *Courtesy from* Philippe Vervier, Sarah Gimet [12] and Gerome Beyries [13].
This grey information (because all documents are not indexed in bibliographic systems) was digitized (PDF to ASCII text) and contents were extracted (text filters), normalized and saved in Open Office documents (using a common template). This new collection of documents was then used as a primary data source by software components: converted in XML (XML DocBook [14]), HTML [15] and analyzed using a custom XML parser. Pertinent fields of information found in documents (year, title, abstract …) were extracted and saved in a CSVM file. *Courtesy from* Jean-Olivier Butron [16].
The following figure shows a Perl CSVM parser that displays CSVM data in a grid widget. The column 'DONNEES' contains links to the corresponding HTML documents (URL|Name of URL), 'ANNEE' stands for year, `'TITLE'` for title and `'RESUME'` for abstract.

*Figure 11. –A simple viewer (wxGrid widget [17]) parsing a CSVM file (a collection of documents).*

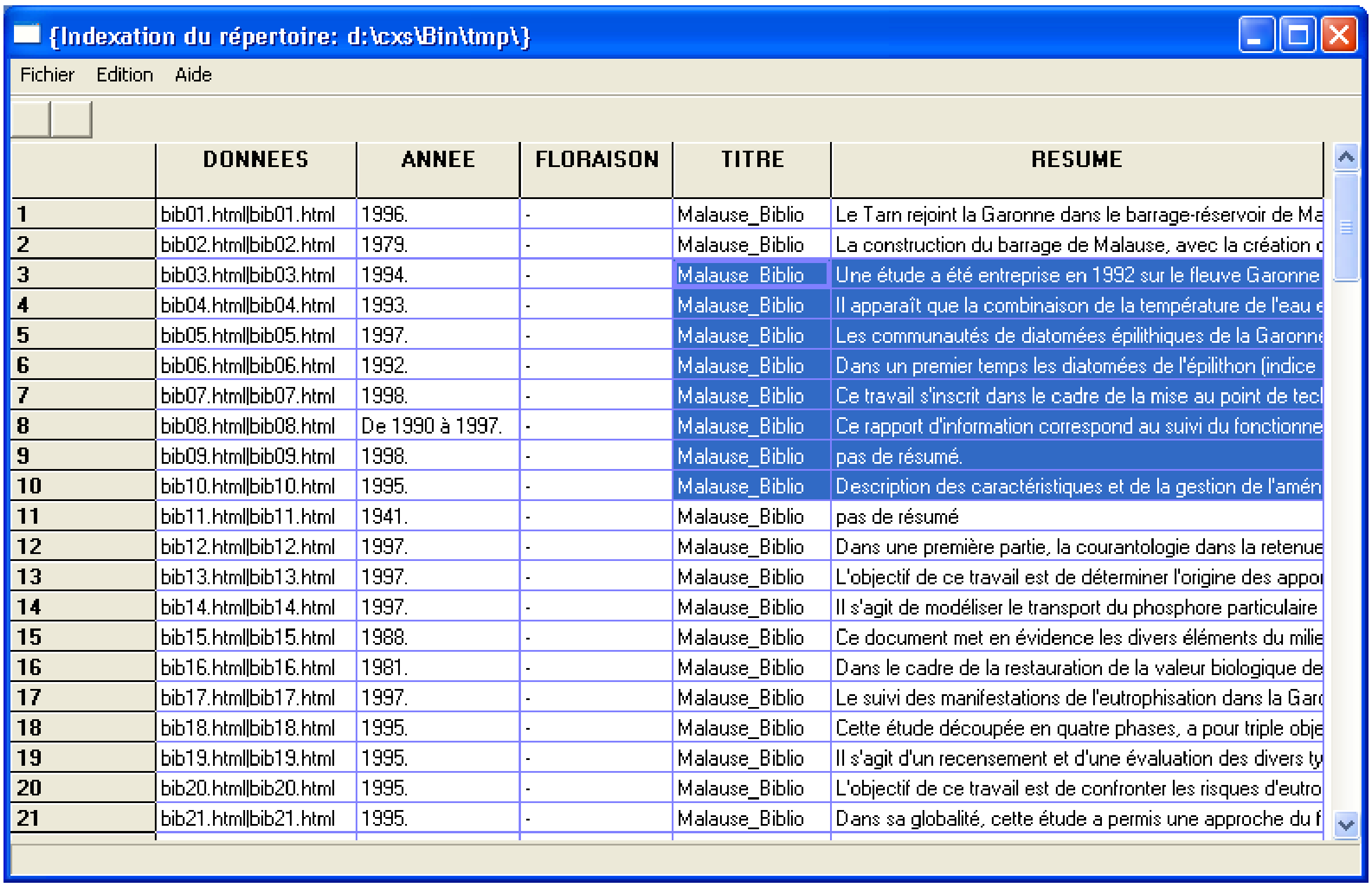

| | DONNEES | ANNEE | FLORAISON | TITRE | RESUME |
|---|---|---|---|---|---|
| 1 | bib01.html\|bib01.html | 1996. | - | Malause_Biblio | Le Tarn rejoint la Garonne dans le barrage-réservoir de Ma |
| 2 | bib02.html\|bib02.html | 1979. | - | Malause_Biblio | La construction du barrage de Malause, avec la création c |
| 3 | bib03.html\|bib03.html | 1994. | - | Malause_Biblio | Une étude a été entreprise en 1992 sur le fleuve Garonne |
| 4 | bib04.html\|bib04.html | 1993. | - | Malause_Biblio | Il apparaît que la combinaison de la température de l'eau e |
| 5 | bib05.html\|bib05.html | 1997. | - | Malause_Biblio | Les communautés de diatomées épilithiques de la Garonne |
| 6 | bib06.html\|bib06.html | 1992. | - | Malause_Biblio | Dans un premier temps les diatomées de l'épilithon (indice |
| 7 | bib07.html\|bib07.html | 1998. | - | Malause_Biblio | Ce travail s'inscrit dans le cadre de la mise au point de tecl |
| 8 | bib08.html\|bib08.html | De 1990 à 1997. | - | Malause_Biblio | Ce rapport d'information correspond au suivi du fonctionne |
| 9 | bib09.html\|bib09.html | 1998. | - | Malause_Biblio | pas de résumé. |
| 10 | bib10.html\|bib10.html | 1995. | - | Malause_Biblio | Description des caractéristiques et de la gestion de l'amén |
| 11 | bib11.html\|bib11.html | 1941. | - | Malause_Biblio | pas de résumé |
| 12 | bib12.html\|bib12.html | 1997. | - | Malause_Biblio | Dans une première partie, la courantologie dans la retenue |
| 13 | bib13.html\|bib13.html | 1997. | - | Malause_Biblio | L'objectif de ce travail est de déterminer l'origine des appoı |
| 14 | bib14.html\|bib14.html | 1997. | - | Malause_Biblio | Il s'agit de modéliser le transport du phosphore particulaire |
| 15 | bib15.html\|bib15.html | 1988. | - | Malause_Biblio | Ce document met en évidence les divers éléments du milie |
| 16 | bib16.html\|bib16.html | 1981. | - | Malause_Biblio | Dans le cadre de la restauration de la valeur biologique de |
| 17 | bib17.html\|bib17.html | 1997. | - | Malause_Biblio | Le suivi des manifestations de l'eutrophisation dans la Garc |
| 18 | bib18.html\|bib18.html | 1995. | - | Malause_Biblio | Cette étude découpée en quatre phases, a pour triple obje |
| 19 | bib19.html\|bib19.html | 1995. | - | Malause_Biblio | Il s'agit d'un recensement et d'une évaluation des divers ty |
| 20 | bib20.html\|bib20.html | 1995. | - | Malause_Biblio | L'objectif de ce travail est de confronter les risques d'eutro |
| 21 | bib21.html\|bib21.html | 1995. | - | Malause_Biblio | Dans sa globalité, cette étude a permis une approche du f |

[12] M. Gerino & coll. Bilan et dynamique de la matière organique et des contaminants au sein d'une discontinuité ex de la retenue de Malause. Restitution des travaux scientifiques du projet de recherche ECOBAG P2 (2006).

[13] G. Beyries. Composants logiciels génériques pour les collections de données. Mémoire ingénieur Ecole des Techniques du Génie Logiciel (2004).

[14] XML DocBook - http://www.docbook.org/

[15] Norman Walsh style sheets - http://nwalsh.com/

[16] F. Rodriguez, G. Beyries, J.O. Butron, C. Blonski. Maîtriser l'information chimique au quotidien: méthodes et composants logiciels à l'interface chimiebiologie. JECB21, Toulouse (2005).

[17] wxWidgets Cross-Platform GUI Library - http://www.wxwidgets.org/

In this case the CSVM file corresponds to the index of this collection and embeds information extracted from documents. At the CSVM level: this is data; at indexed files (XML files converted in HTML files) level: this is metadata. All the different tasks were processed automatically without data input 'at hand' by an operator. A lot of conversion between files (Raw text, HTML, XML, XML-Docbook, HTML) were also configured and driven by CSVM files. The following figure shows a CSVM file used to drive changes in the CSVM index itself (each row is a filter applied to the index).

*Figure 12. – Some rows of a CSVM files filled with rules for a text filter.*

```
STR	{HTML: }	-	-	-	-	-	-	-	-	-
STR	XML index:	Indexation du répertoire:	-	-	-	-	-	-	-	-
STR	-dbk.html	.html	-	-	-	-	-	-	-	-
STR	-dbk.htm	.html	-	-	-	-	-	-	-	-
STR	_TABAbstract	_TABRESUME	-	-	-	-	-	-	-	-
STR	_TABabstract	_TABRESUME	-	-	-	-	-	-	-	-
STR	_TABTitle	_TABTITRE	-	-	-	-	-	-	-	-
STR	_TABtitle	_TABTITRE	-	-	-	-	-	-	-	-
STR	_TABfile	_TABDONNEES	-	-	-	-	-	-	-	-
STR	_TABFile	_TABDONNEES	-	-	-	-	-	-	-	-
STR	hashtable2{ANNEE}	ANNEE	-	-	-	-	-	-	-	-
STR	hashtable2{FLORAISON}	FLORAISON	-	-	-	-	-	-	-	-

#TITLE	Change some values in CSVM indexes
#HEADER	CODE	TAG	TAG	TAG	TAG	TAG	TAG	TAG	TAG	TAG	TAG
#TYPE	TEXT	TEXT	TEXT	TEXT	TEXT	TEXT	TEXT	TEXT	TEXT	TEXT	TEXT
#WIDTH	50	50	50	50	50	50	50	50	50	50	50
```

The first column is used to store the type of filter using a keyword. The keyword `STR` is used to specify a string replacement of the value stored in second column by the value stored by the third column. More complex filters that '`STR`' exist and need to use fully the 10 columns. Then the CSVM index was integrated with a JavaScript framework in order to provide a view of the collection dedicated to end users. Buttons at the top of dynamic table are used to sort data of CSVM index by a given column.

*Figure 13. – Same data (as shown in Figure 10) parsed using a JavaScript list component and displayed as item list (bibliographic style) rather than columns. Data are sorted by year (ANNEE) and displayed in the order (ANNEE, DONNEES, FLORAISON, TITRE, RESUME).*

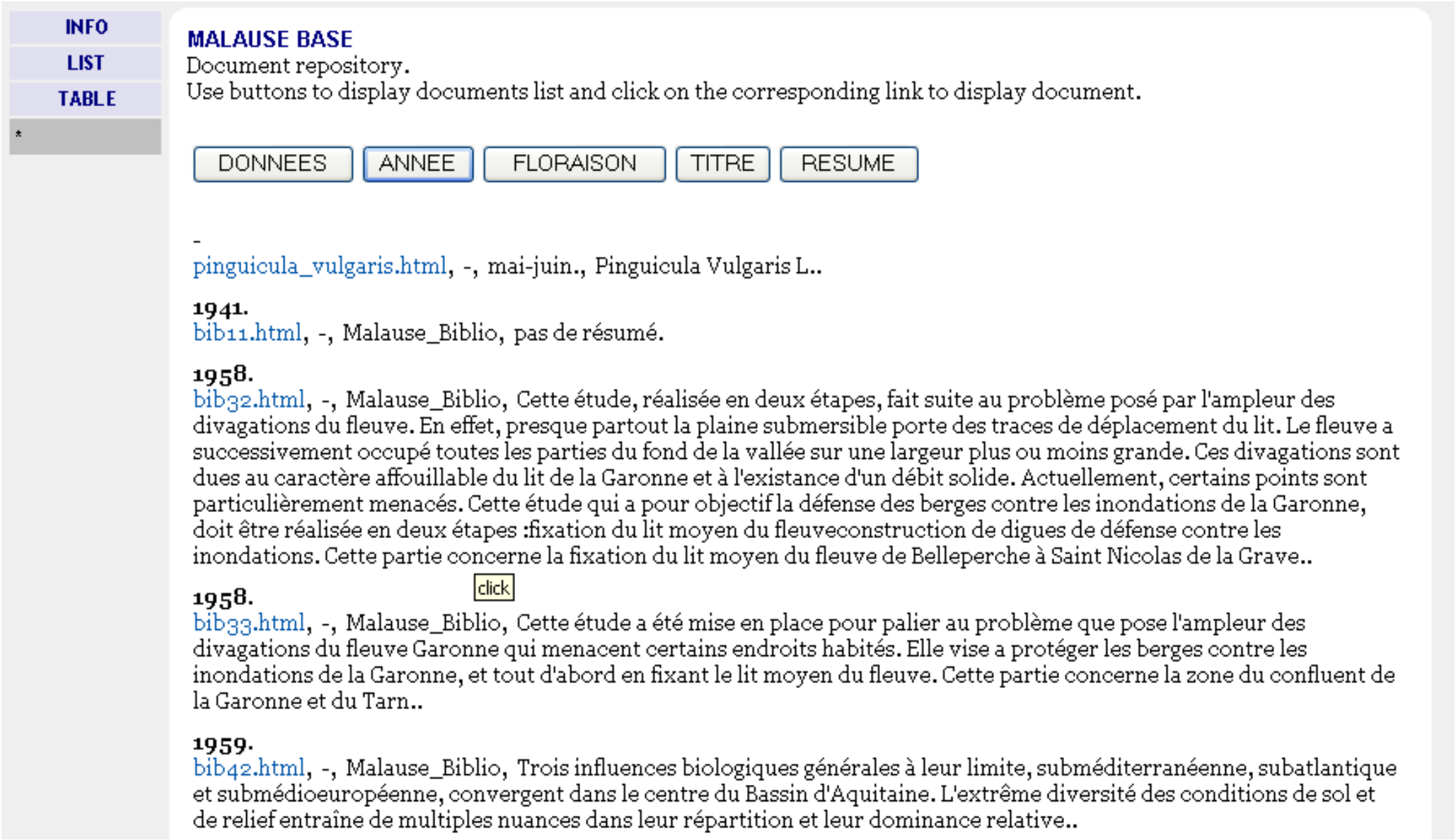

When the publishing chain works, all to do is to write the CSVM file and to process the chain. Edition of the CSVM index can be done, manually (text editor), using a GUI, automatically by an indexer component). In this case the documents were stored in HTML (after processing) or native (hyperlink to the corresponding file or image) formats.

The following figure shows final results for this kind of publishing chain: *14-A)* the case of small organic molecules (ligands, inhibitors) interacting with some enzymes (GlcCerase, beta-glucosidase) and *14-B)* the case of enzymes itself (HAT).

*Figure 14-A. – Chemical collection of ligands taken from the Protein Data Bank [18].*

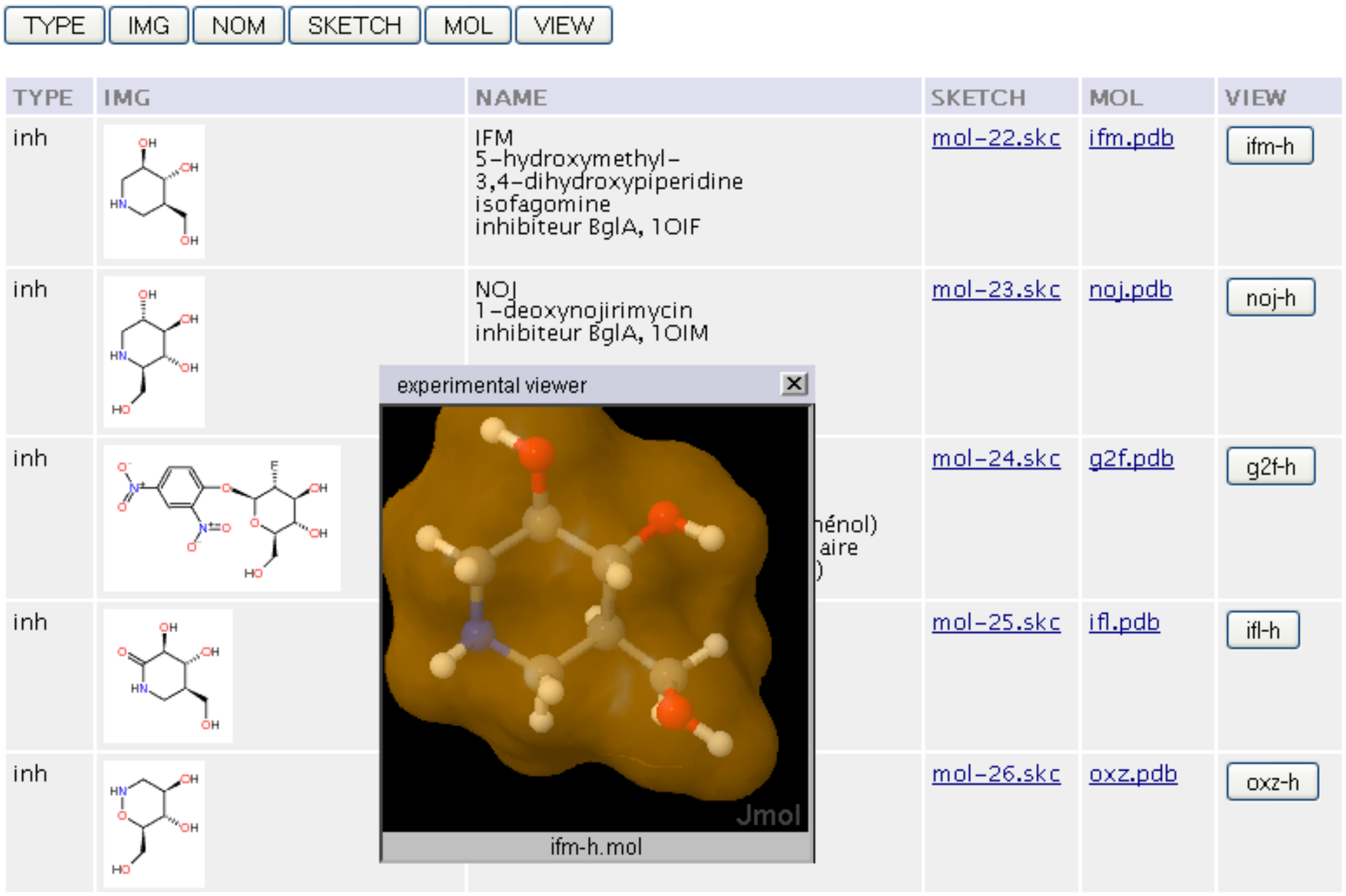

*Figure 14-B. – Collection of macro-molecules (enzymes) structures taken from the Protein Data Bank.*

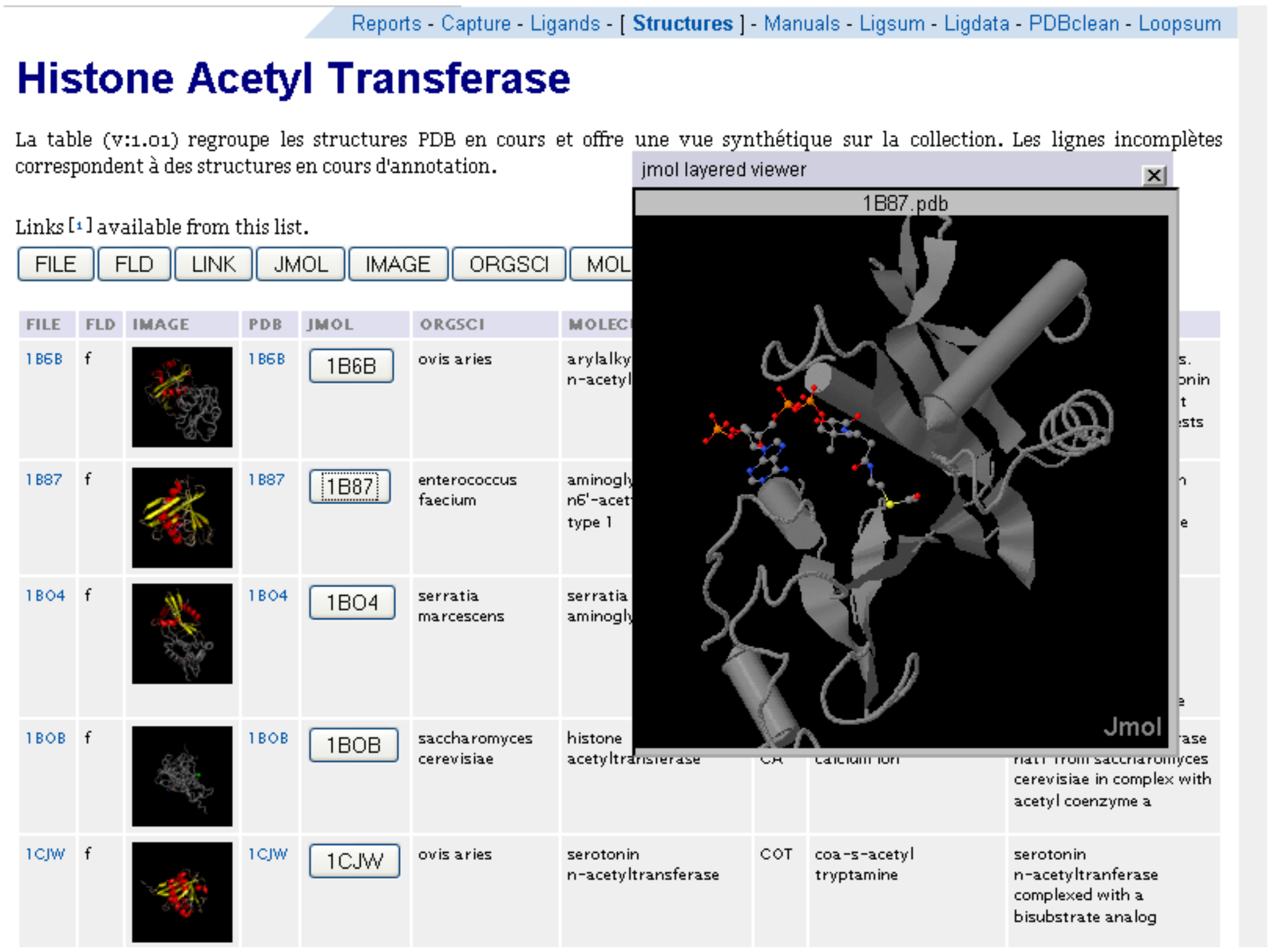

---

[18] RCSB Protein Data Bank – http://www.pdb.org/ a main source for crystallographic structures of proteins.

*Courtesy from* Marjorie Catala [19] ; Casimir Blonski, Christian Lherbet and Cecile Baudoin-Dehoux [20] [21] .

The previous figure (14-A) shows the case of a chemical collection (small organic molecules). The collection mixes images ('IMG' column), some information (TYPE, NOM), 2D (SKETCH) and native 3D (MOL) structures and action buttons (VIEW) launching an external viewer [22] embedded (HTML layer) in a floating window. In the case of Figure 14-B, the interface of the molecular collection (macromolecules) is multimodal and could launch local/hyperlinks (`'FILE'/'LINK'` columns), PyMol [23] generated images (`'IMAGE'`) of structure, a viewer (`'JMOL'`) as other information.
From a CSVM point of view (RAW data) the samples of Figure 13-14-15 are of the same complexity despite the kind of objects displayed. All is taken in account by a different JavaScript framework: the CSVM data were converted in JavaScript tables and were processed by dedicated JavaScript components. The Following Code shows the contents of this kind of JavaScript table and the first row for the molecular structure displayed in the case of Figure 14-A:

*Code 3. – A JavaScript table descending from a CSVM file.*

```
// GXMLPARSER interface, download these data before,
// and use GXMLDISPLAY for use whith browsers

function jmol(mol,button)
{       var src = "<INPUT onclick=" + '"' + "loadwindow('jmol.html',250,287,'" + mol +
"');" + '"' + "type=button value='" + button + "'>";
        return(src);
}

var flags_array = new Array ("TYPE", "SECTOR", "IMG", "NAME", "SKETCH", "MOL", "VIEW");
var flags_n=7;
var getf_array = new Array ("TYPE", "IMG", "NAME", "SKETCH", "MOL", "VIEW");
var getf_n=6;
var data_c=7;

var data_array = new Array();
var data_r=0;
...
data_array[data_r++] = new Array ("inh", "glccerase", "<IMG SRC='sketch/mol-22.png'>",
"IFM<br>5-hydroxymethyl-3,4-dihydroxypiperidine<br>isofagomine<br>inhibiteur BglA, 1OIF",
"<A HREF='sketch/mol-22.skc'>mol-22.skc</A>", "<A HREF='lig/ifm.pdb'>ifm.pdb</A>",
jmol('ifm-h.mol','ifm-h'));
...
```

*It is easy to transform a CSVM file in another tabular data (i.e. JavaScript, CSV, XML tables) format because all that is needed to do the transformation is embedded in the CSVM file itself.*

The corresponding index CSVM files can be edited manually by users (as they were using spreadsheet tables) or automatically generated, but the CSVM mechanism has the advantage of providing a lot of features for further processing and transformations.

19 M. Catala - Etude structurale et arrimage moléculaire (Docking) *in silico* de prodrogues dans la ß-glucocérébrosidase. Rapport M1BBT (2006) Université Paul Sabatier.

20 S. Gau, F. Rodriguez, C. Lherbet, C. Menendez, M. Baltas. Approches interdisciplinaires pour la conception rationnelle d'inhibiteurs. RECOB 13, Aussois (2010).

21 New potent bisubstrate inhibitors of histone acetyltransferase p300: design, synthesis and biological evaluation. F.H.A. Kwie, M. Briet, D. Soupaya, P. Hoffmann, M. Maturano, F. Rodriguez, C. Blonski, C. Lherbet, C. Baudoin-Dehoux. *Chem Biol Drug Des.* (2011) 77:1, 86-92.

22 JMOL: An open-source Java viewer for chemical structures in 3S – http://jmol.sourceforge.net/

23 PyMol – http://www.pymol.org/

## 4.2. Environmental tabular data

This sample shows a way to dig inside files when the indexed files are also CSVM files. An index was also created but each entry was linked to a CSVM file rather than a text document (Open Office, XML, HTML, PDF …) or a data source (data file, image, hyperlinks).
In this case, the CSVM data (table of measurements, not index) were written manually using a standard spreadsheet. *Courtesy from* Magali Gerino, Maya Lauriol, Sabine Sauvage [24] and Gerome Beyries [25].

As in previous case, end users can download corresponding CSVM files using an hyperlink in index display, but also they can view (after another CSVM conversion in JavaScript tables of each leaf files) directly inside these files using another JavaScript framework [26], as shown by the following figure:

*Figure 15. – A dynamic view of a CSVM file after its conversion (JavaScript table, columns at right not shown).*

[malause1.js] **Malause station 1**

| date | heure | numero station | coordonnees gps | profondeur | methode de mesure | te |
|---|---|---|---|---|---|---|
| 2003/07/07 | 12:00 | 1 | 0346600-4884900 | 0 | sonde manuel | |
| 2003/07/07 | 12:00 | 1 | 0346600-4884900 | 0.25 | sonde manuel | |
| 2003/07/07 | 12:00 | 1 | 0346600-4884900 | 1.75 | sonde manuel | |
| 2003/07/07 | 12:00 | 1 | 0346600-4884900 | 2 | sonde manuel | |
| 2003/07/07 | 12:00 | 1 | 0346600-4884900 | 2.25 | sonde manuel | |
| 2003/07/07 | 12:00 | 1 | 0346600-4884900 | 2.5 | sonde manuel | |
| 2003/07/07 | 12:00 | 1 | 0346600-4884900 | 2.75 | sonde manuel | |
| 2003/07/07 | 12:00 | 1 | 0346600-4884900 | 3 | sonde manuel | |

Showing No. 1 to 13 of 13

|< < Query * > >|

In this table the `#WIDTH` information is used for column's widths and `#TYPE` values are injected as column's names in the JavaScript tables.

*CSVM can be used as well as for index and for tables. This approach has a great potential for complex processing of files because one parser and one format are used independently of tabular objects and information.*

[24] M. Gerino & coll. Bilan et dynamique de la matière organique et des contaminants au sein d'une discontinuité ex de la retenue de Malause. Restitution des travaux scientifiques du projet de recherche ECOBAG P2 (2006).

[25] G. Beyries, F. Rodriguez. Quels outils informatiques pour les collections de données ? Restitution du Programme ECOBAG P1, Agen (2004).

[26] A modified version of Dieter Bunger's (http://www.infovation.de) Cross Browser Dynamic HTML Tables code - http://www.insidedhtml.com/tips/databinding/ts06/paper/dhtml_table.htm.

## 4.3. Working with RAW tables

Typically a case encountered in collaborative projects, in which the aggregation of RAW data from different scientific fields (and different data types by definition) is expected. The CSVM concept makes easy to compute intersection or union of tables which could be often useful. These operations are possible even if the tables cover the 1) same, 2) a common core, and 3) different sets of data. In Python CSVM toolkit the union or intersection of CSVM tables use csvm_ptr() objects :

```
print "\n*** Test completely different CSVM files"
c1 = csvm_ptr()
c1 = csvm_ptr_read_extended_csvm(c1,"test/test1.csvm","\t")
c2 = csvm_ptr()
c2 = csvm_ptr_read_extended_csvm(c2,"test/test2.csvm","\t")
```

The two CSVM files were loaded in two csvm_ptr() objects **c1** and **c2**, now the intersection of the two tables can take place, the result is a new csvm_ptr() named **r** :

```
print "=> Compute INTERSECTION"
r = csvm_ptr_intersect(c1, c2)
if (r != None):
    r.csvm_ptr_dump(0,0)
    r.csvm_ptr_clear()
else:
    print "No data found"
```

The union of tables is coded as same, and after operation all data structures are destroyed [27].

```
print "\n=> compute UNION"
r = csvm_ptr_union(c1, c2)
r.csvm_ptr_dump(0,0)
r.csvm_ptr_clear()
c1.csvm_ptr_clear()
c2.csvm_ptr_clear()
print "\n*** Test csvmutil done."
```

Notice that after (and sometimes before) union of tables with different (names and types) columns, the RAW data can be sparse, typically a big number of columns, without data for some (or nearly all) rows. If same columns are found in **c1** and **c2** tables, the union mechanism preserves data because data blocks (rows) are added successively and not interleaved.
For an attentive reader, just remark something on Figure 13: the presence of a file named *pinguicula_vulgaris*. This file contains information about a carnivorous plant (butterworts) which seems not to be included in the same category than the others (bibliographic digests about stations around Malause: *malause_biblio*): this is the result of union of tables from different CSVM files.
In the case of *pinguicula* a year was not extracted or assigned from data file, and the resulting cell in final CSVM file is marked by the empty character (`'-'` in this case), the CSVM *puinguicula* branch has not this column. The last column is an excerpt of the [abstract/beginning of the text] found in files. The *pingucula* type CSVM files have not this kind f information. But, the text recognized as title (not filename) inside the document was assigned to a particular column. This column was later and prior to union of tables set as equivalent to the `'RESUME'` column of *malause_biblio* branch. So this column is filled in the resulting CSVM file that was processed in the dataflow. Another mechanism was also used: adding [28] a column `'FLORAISON'` (flowering date) not found in the *malause_biblio* branch and obviously found in *pinguicula* branch.

*This shows how it is possible to manipulate and aggregate RAW data easily in CSVM space without any data modeling*

---

[27] Directly taken from the *build.parsers.csvm.csvmutil* Python module (unitary test) of the CSVM toolkit.
[28] Python CSVM toolkit makes it possible using one line of code.

## 5. From collections to calculations: the case of time series

The case of enzyme kinetics is interesting because the CSVM format can be used in different ways: to store *i)* data (time series); *ii)* parameters for calculation (modeling, curve fitting); *iii)* results (best fit, residuals, parameter's values) and sometimes *iv)* the model itself. These steps can be found in a lot of other scientific fields (i.e. Physics, Chemistry) in which modeling is used as a core technique.
Enzyme kinetics measurements or calculations are typical of that we call short time series (less than 5000 rows, typical: 10 to 1000). This kind of data-island is widely found in a lot of biological, biochemical, chemical measurements, *courtesy from* Sabine Gavalda [29] [30] and Casimir Blonski.
This example, taken from real laboratory work, is typical for large sets of kinetics, a set of 800 files (RAW data) were collected and processed [31] without using any database system.

### 5.1. A canonical format

The corresponding data is a molar (or equivalent measurements: i.e. values of optical densities taken with a spectrophotometer) concentration ([S] decreasing or [P] increasing of a Substrate (P) or a Product (P) depending from time (the first column tagged '`Time (s)`'). Each column at right of time, is an evolution of [P] for a particular molar concentration of an Inhibitor, giving a table of *[P] = f(t, [I])* which can be modeled later to give parameters of an underlying kinetic model.
A typical example is given in the following figure with the first column devoted to time base and the others at right to different experiments (increase of Product corresponding to a particular value of inhibitor concentration).

*Figure 16-A. – A partial view of kinetics CSVM data, a lot of rows are not shown (marked by '…' characters).*

```
0.000   -0.0001  -0.0007  0.0015   -0.0002  0.0003   0.0010      -0.0013  -0.0017  0.0008   0.0009   -0.0017  0.0002   0.0010   -0.0004
0.010   0.0001   -0.0013  0.0004   -0.0004  -0.0011  -0.0000040  -0.0014  -0.0015  0.0001   0.0006   -0.0018  0.0005   0.0004   -0.0010
0.020   0.0001   -0.0009  0.0014   -0.0003  -0.0004  0.0005      -0.0013  -0.0014  -0.0004  0.0021   -0.0025  -0.0003  0.0008   -0.0003
...
6.970   0.4313   0.4479   -        -        -        -           -        -        -        -        -        -        -        -
6.980   0.4324   0.4470   -        -        -        -           -        -        -        -        -        -        -        -
6.990   0.4328   0.4484   -        -        -        -           -        -        -        -        -        -        -        -

#TITLE  csvm\sab150403.csvm (CSVM) from [reports\sab150403.txt]
#HEADER Time (s)     A1       A2       A3       A4       A5       A6         A7       Set-8    Set-9    Set-10   Set-11   Set-12   Set-13   Set-14
#TYPE   NUMERIC      NUMERIC  NUMERIC  NUMERIC  NUMERIC  NUMERIC  NUMERIC    NUMERIC  NUMERIC  NUMERIC  NUMERIC  NUMERIC  NUMERIC  NUMERIC  NUMERIC
#WIDTH  50           50       50       50       50       50       50         50       50       50       50       50       50       50       50
#META
```

This primary CSVM file can be generated *de novo* by users or acquisition system, but it can also result from another data format specific of a method or an apparatus. This was the case for this sample and the original format is shown in the following figure:

*Figure 16-B. – Original format issued from a spectrophotometer..*

```
  X       Y       X       Y       X       Y       X       Y
----------------------------------------------------------------
 0.000   0.1112  0.050   0.1111  0.100   0.1111  0.150   0.1113
 0.200   0.1116  0.250   0.0875  0.300   0.0514  0.350   0.0378
 0.400   0.0696  0.450   0.1014  0.500   0.1126  0.550   0.1135
 0.600   0.1145  0.650   0.1155  0.700   0.1164  0.750   0.1173
 0.800   0.1181  0.850   0.1188  0.900   0.1194  0.950   0.1202
...
```

This example show one interest of the CSVM format: to be used as a canonical data format for RAW tabular data. This permits to use data independently of the data source and in a better way than pure CSV files that cannot embed metadata and annotations.

29 S. Gavalda. L'Enolase de *Trypanosoma brucei* : Synthèse et étude du mode d'action d'inhibiteurs. Thèse de doctorat de l'Université Paul Sabatier Toulouse III (2005).

30 S. Gavalda, F. Rodriguez, G. Beyries, C. Blonski. Software design and components for enzymology,. 21ème CBSO, Ax-Les-Thermes (2006).

31 S. Gavalda, F. Rodriguez, G. Beyries, C. Blonski .Méthodes et composants logiciels pour l'interface chimie-biologie. RECOB11 Aussois (2006).

In the CSVM file shown in Figure 16-A, the amount of information seems to be low: the values of inhibitor concentration (#HEADER field) are tagged as `A1, A2, … Set-14` which is not very explicit if the enzymologist wants to reuse the data later. In this case, the main information is the number of file: 150403 (date of experiment) stored in #TITLE field, this record number give access to another file with contains full information about experimental conditions.
This approach was used because we were not sure, at this time, if it was pertinent to store all parameters of the experiment with the data in the same CSVM file. It was possible to use the `#META` field for that, in another way the insertion of annotations in the CSVM table could be used [32].
Notice also that some of the last rows of this file are not filled with data. This is normal, all kinetics are not recorded with same duration (but with same time origin and intervals). The CSVM convention lets you to mark these empty positions by a particular character (in this case `'-'`) or not.

## 5.2. Data extraction from RAW values

Very often calculation programs cannot handle multi-column files (not only CSVM) but uses a two-column file: time and a variation of a given quantity (here the product of reaction, for a particular value of inhibitor concentration). The following figure shows this kind of file with: a number of experimental points in the first row, followed by two columns using spaces as delimiters, one for time, the other for a biochemical quantity, and at last a block (rows beginning by a `#` character) for some metadata.

*Figure 17. – A two-column data file specific of a given software.*

```
691
0          0
0.05       0.67669
0.1        1.42857
0.15       2.18045
0.2        2.85714
...
34.383     281.50376
34.433     281.8797
34.483     282.25564

# LPZ file . lpz\0234-1.lpz
# Npts ..... 691
...
# Colx no .. 0
# Header x . Time
# Coly no .. 1
# Header y . B
...
```

In this case, we need to use software components to extract columns from the primary CSVM file, make this kind of dataset: {`Time (s); A1`}, {`Time (s); A2`} … {`Time(s); Set-14`} and save it using the corresponding format. With CSVM it is easy to split the multi-column file (one time column, $i=1..n$ columns $[P]_i$) into $n$ CSVM files (each on with time and $[P]_i$). Column extractors are included in Python toolkit, a Python list is returned, depending on a search string applied on #HEADER values:

*Code 4. – Simple CSVM column extractor in Python language.*

```
print "*** Extract columns on the value of headers"
print "-> the column named 'fichier_mol' in strict mode (equal)"
ls = csvm_ptr_getcol(c, 'fichier_mol', 1)
print "found %d column in CSVM stream" % (ls[0])
print ls[1]
print "-> the columns in which string 'no_' is found (include)"
ls = csvm_ptr_getcol(c, 'no_', 0)
print "found %d columns in CSVM stream" % (ls[0])
```

32 And easy to do: if the first character of a row is a # not followed by a reserved keyword (TITLE, META, HEADER, TYPE or WIDTH) the row is not read by parsers and is available to store annotations, this kind of annotation provides a high level of information (but not immediately available for processing) for the related table.

For large sets it is also possible to make queries based on all parts of CSVM (metadata and data) following the columns or rows. The following output shows on a simple string (green) matrix (the data block of a CSVM Python class) how to make a search or a combined search (AND, OR) on columns or rows:

*Code 5. – Example of using a CSVM table for a file catalog (this output is a dump of a CSVM Python class).*

```
*** Test EQ column queries samples
uses matrix :  [['PDB', '3.24', 'AB4'], ['-', '1.0', '46'], ['PDB', '1.01', '4']]
query_col_eq(matrix, '4', 2, 1) => strict : rows  [2]
query_col_eq(matrix, '4', 2, 0) => inc : rows  [0, 1, 2]
query_col_eqs(matrix, 'PDB -', 'and', ' ', 0) => strict/and : rows  []
query_col_eqs(matrix, 'PDB -', 'or', ' ', 0) => strict/or : rows  [0, 2, 1]
query_col_eqs(matrix, '4 46', 'and', ' ', 2) => strict/and : rows  []
query_col_eqsv(matrix, ['4','46'], 'or', 2) => strict/or : rows  [2, 1]
query_col_eqsv(matrix, ['4','46'], 'or', 2, 0) => inc/or : rows  [0, 1, 2]

*** Test NOT.EQ column queries samples
uses matrix :  [['PDB', '3.24', 'AB4'], ['-', '1.0', '46'], ['PDB', '1.01', '4']]
query_col_not_eq(matrix, '4', 2) => strict : rows  [0, 1]
query_col_not_eq(matrix, '4', 0, 0) => inc : rows  [0, 1, 2]
query_col_not_eqs(matrix, '4 LIG', 'or', ' ', 0, 0) => inc : rows  [0, 1, 2]
query_col_not_eqs(matrix, 'B 6', 'and', ' ', 2, 0) => inc : rows  [2]

*** Test EQ row queries samples
get the first row of matrix :  ['PDB', '3.24', 'AB4']
query_row_eq(row, '4') => strict : columns  []
query_row_eq(row, '4', 0) => inc : columns  [1, 2]
query_row_eqsv(row, ['4','PDB'], 'or', 0) => inc/or : columns  [1, 2, 0]

*** Test NOT.EQ row queries samples
get the first row of matrix :  ['PDB', '3.24', 'AB4']
query_row_not_eq(row, 'PDB') => strict : columns  [1, 2]
query_row_not_eqs(row, 'B 4', 'and', ' ', 0) => inc/and : columns  []
query_row_not_eqs(row, 'B 4', 'or', ' ', 0) => inc/or : columns  [1, 0]
query_row_not_eqs(row, 'LIG -', 'or', ' ', 0) => inc/or : columns  [0, 1, 2]
```

The results (red) of each tested functions (blue) are enclosed in brackets at right of lines. In strict mode the search string (or the list of search strings) must be found as a plain value of a cell, if not strict (in mode), the search string can be also a substring of a cell value.
With this minimal toolkit it is easy to extract columns or rows and to make files at a given format usable by software that cannot handle multi-column CSVM files. CSVM files and processors can simplify greatly the handling of complex dataflows.

## 5.3. Mixed time series and information

In this case, two column curves were extracted from the CSVM file (the primary data source) shown in Figure 16-A. Each extracted two-column files is formerly a curve $[P]=f(t)$ and it can be used for calculation (curve fit). But what about for the results ? In the following sample, not only the results of minimization (data, model key, residuals) were stored in one CSVM file but also all parameters used for calculation (initial values, solution, confidence intervals …) by adding new columns at right of data.
The following output is a typical result after a calculation. The four first columns code for time series: column 1 for time (`X0`), column 2 for product [P] concentration (`Y0`), column 3 for curve fit using model C23 (`Y0 CALC`) and column 4 for residuals (`RES`) between experimental (Y0) and calculated (Y0 CALC) data.

*Figure 18. – Results of a curve fitting calculation enclosed in a CSVM file.*

```
0       0         0         0          LFILE    A020304\LPZ\0234-17.LPZ
.05     .97744    .6794984  .2979416   PROG     Model C23
.1      1.8797    1.345653  .5340471   VER      v:(1.0)
.15     2.85714   1.998803  .8583375   NPTS     687
.2      3.83459   2.639278  1.195312   NPAR     3
.25     4.3609    3.267402  1.093498   NVAL     1
.3      4.66165   3.883489  .7781613   F        182.0163
.35     5.11278   4.487844  .6249361   ET       .5158544
.4      5.86466   5.080766  .7838941   PNUM     1
.45     6.61654   5.662545  .9539948   PNAME    k
.5      7.21805   6.233466  .9845843   PSOL     .5151757
```

```
.55     7.81955     6.793804    1.025746       PMIN     .5145575
.6      8.42105     7.343827    1.077223       PMAX     .5157939
.65     8.79699     7.883799    .9131913       PNUM     2
.7      9.24812     8.413976    .8341446       PNAME    Vo
.75     9.62406     8.934606    .6894541       PSOL     8.571785
.8      10.07519    9.445931    .6292582       PMIN     8.558927
.85     10.67669    9.94819     .7285004       PMAX     8.584643
.9      11.2782     10.44161    .8365898       PNUM     3
.95     11.65414    10.92642    .7277212       PNAME    Vs
1       11.8797     11.40284    .4768639       PSOL     3.095204
1.05    12.10526    11.87107    .2341881       PMIN     3.092419
1.1     12.85714    12.33134    .5258017       PMAX     3.09799
1.15    13.60902    12.78383    .8251867       CNUM     1
1.2     14.13534    13.22876    .9065809       CNAME    Quality
1.25    14.58647    13.6663     .920166        CVAL     991
1.3     15.03759    14.09666    .9409323       END      -
1.45    16.01504    15.34638    .6686592       -        -
1.5     16.39098    15.74975    .6412268       -        -
1.55    16.69173    16.1468     .5449276       -        -
 ...
34.183  125.4135    126.4381    -1.024567      -       -
34.233  125.5639    126.5929    -1.028954      -       -
34.283  125.8647    126.7476    -.8829575      -       -
#
#TITLE      / Model C23
#HEADER     X0         Y0         Y0 calc     RES         KEY        VALUE
#TYPE       NUMERIC    NUMERIC    NUMERIC     NUMERIC     TEXT       NUMERIC
#WIDTH      50         50         50          50          50         50
#META       CSVM Result / VA04A QB BUILD
```

The two last columns are used to store calculation parameters and some results as a *key-values* scheme. This information are known as columns 'KEY', 'VALUE' and of type TEXT and NUMERIC and include primary data file (LFILE), program used (PROG), number of experimental points (NPTS), calculation values such as standard deviation (ET) and solution (PSOL) of model's parameters (PNAME) with confidence intervals (PMIN, PMAX), or other post processing calculations (Quality).
The *KEY-VALUE* block is closed by a KEY='END', after this point the 'KEY' and 'VALUE' columns are filled with blank chars ('-' in this case).

## 5.4. Combine different layers of data and calculations

This kind of file is considered a secondary data source, and these files could be aggregated to do other calculations. But, in fact, the most frequent use of secondary data is to pick some results inside (typically a solution value for a given parameter to be optimized). In this case, the value (PSOL=0.5151757) of an apparent kinetic constant *k* (PNUM=1, PNAME=k) was extracted, and stored in a new CSVM file: the tertiary data source.
More precisely, this file was used to store the variation of *k* value (secondary data) with concentrations of [S] or [I] (substrate and inhibitor) stored in the primary data source (CSVM file shown in Figure 16). An example of tertiary data source is given in the Figure 19 :

*Figure 19. – Aggregated results in a CSVM file and used as a new data source in the next calculation layer. Some columns are not shown and are marked by '...' symbols.*

```
A020304  0234-1.LPZ   Model C23   691    3     1     186.8595 .5211507  1  k ...
A020304  0234-10.LPZ  Model C23   686    3     1     224.6174 .5734709  1  k ...
A020304  0234-11.LPZ  Model C23   691    3     1     1373.348 1.41285   1  k ...
A020304  0234-12.LPZ  Model C23   690    3     1     152.6696 .4714091  1  k ...
A020304  0234-13.LPZ  Model C23   684    3     1     509.5747 .8650284  1  k ...
...
TB310304 3134-8.LPZ   Model C23   623    3     1     516.7548 .9129487  1  k ...
TB310304 3134-9.LPZ   Model C23   649    3     1     682.8931 1.028159  1  k ...

#TITLE   Results in [model_c23\][CSV]
#HEADER  LFILE  PROG  NPTS  NPAR   NVAL  F     ET    PNUM   PNAME ...
#TYPE    UNDEF  UNDEF UNDEF UNDEF  UNDEF UNDEF UNDEF UNDEF  UNDEF ...
#WIDTH   50     50    50    50     50    50    50    50     50    ...
```

The tertiary data source was then used to make other calculations in order to compute an inhibition constant ($K_I$ or $K_I$*) significant of the affinity of a given inhibitor relative to this enzyme and that is used to rank the efficiency of different inhibitors.
The details of equations used in this software pipe are given in Annex-1 with the same color conventions than in this section.

## 5.5. Data driven parallelism

This previous sample shows how CSVM could be used to follow closely data parallelism, the Figure 19 generalize this kind of concept:

*Figure 20. – Data parallelism in the case of this Enzyme kinetics processing.*

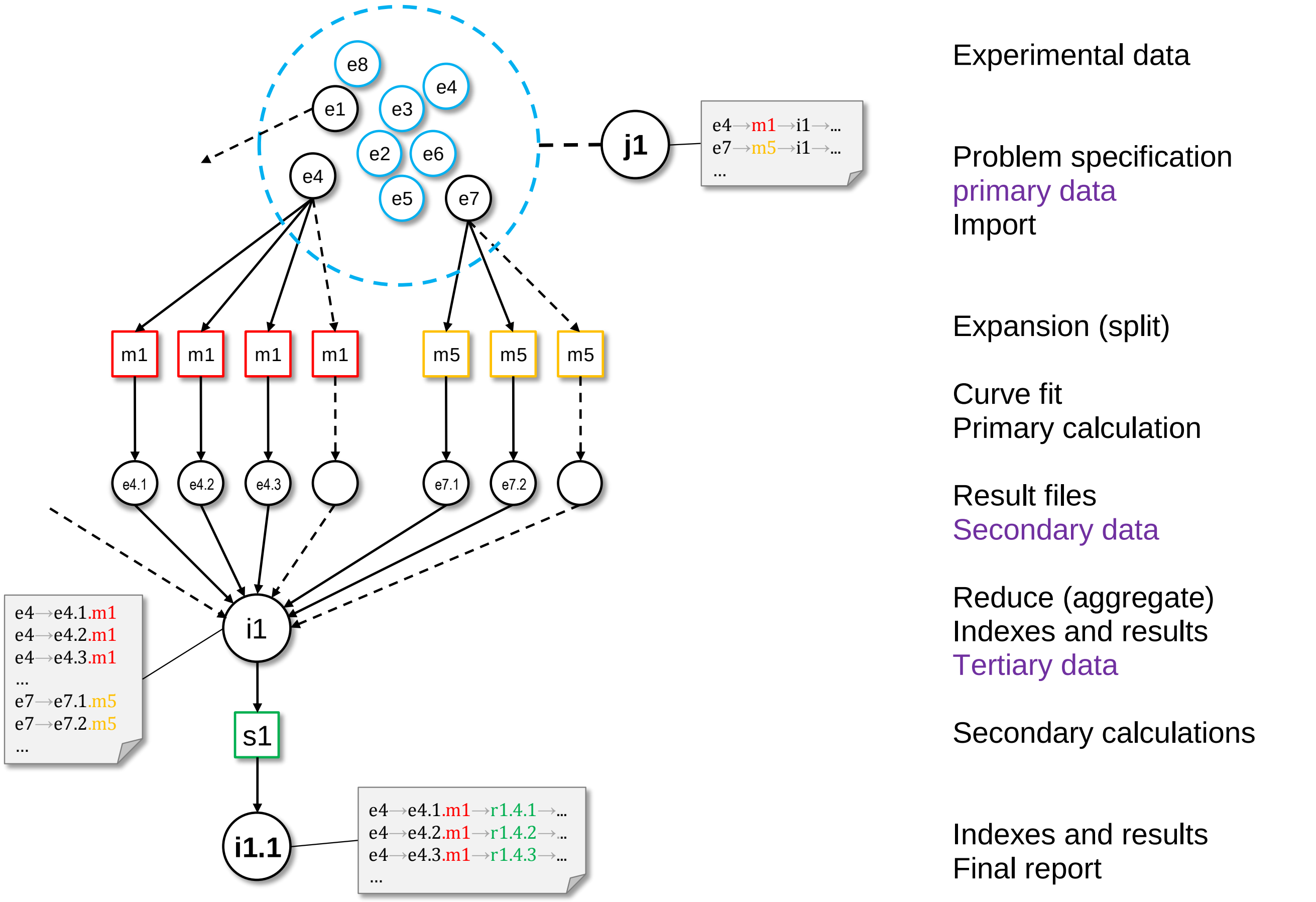

The files in set **j1** are native files (apparatus or other formats). These files (**e4**, **e7** …) are converted to the canonical format in order to create a CSVM data space (e1, e2, e3 … e8). For each file of this space the *n* data sets (two-column sets *time* vs. *quantity*) are extracted, and *n* suitable files for calculation (CSVM or not depending on the software used) are created (i.e. m1 or m5 squares).
After calculation the results are stored in *n* CSVM files mixing original data and results (i.e. e4.1, e4.2, e4.3 … for columns m1, m2, m3 … of file e4).
Some data inside these files are then extracted and aggregated inside a new data source **i1**. This source can be used as a report or for a new calculation run. In this case a new file can be created or columns can be added to **i1** that gives a modified **i1.1**
All calculations are independent and could be attributed to a pool of nodes/threads depending on the parallel model used. We can use CSVM as support for a data driven parallelism and it could be possible to implement an abstraction layer upon CSVM to make operations such as: split, scatter/bcast, reduce … (typically MPI-like operations on data at different calculation steps) in order to implement a model of high throughput enzymology.

## 5.6. From calculations to reports

In the previous sections we have shown that CSVM is well suited to support publication chains. This also the case here, the final (i.e. **i1.1**) or intermediate reports (**i1**) can be used as input for a new dataflow (calculation or data exposition). The following figure shows a proof of concept for a report layer at the level **i1** that includes a JavaScript framework and software components for extracting data (i.e. make graphs) from the CSVM files.

*Figure 21. – Output of a publishing chain using aggregated data from calculations.*

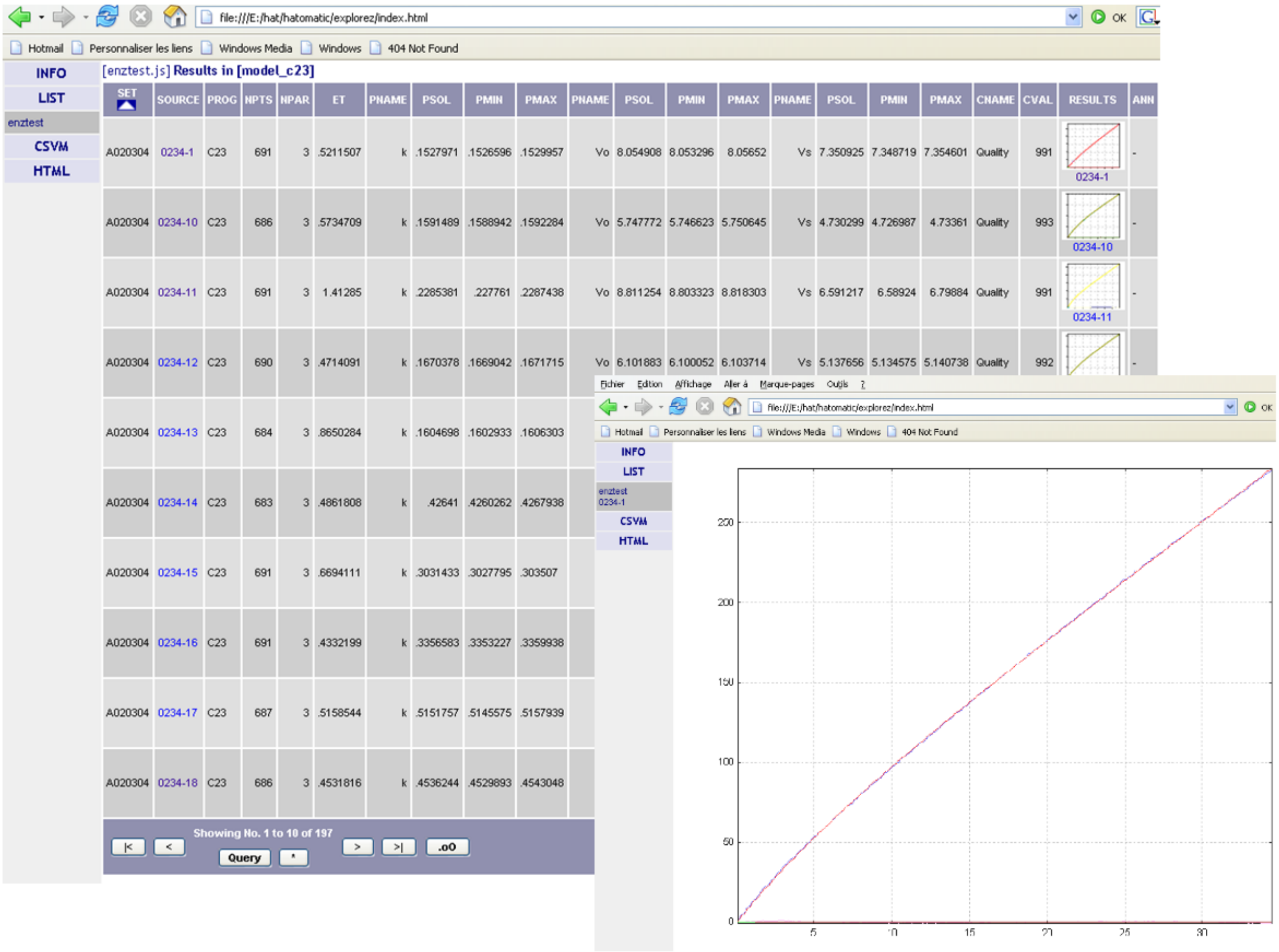

## 5.7. Including models in the scheme

Calculations uses a phenomenological model for data fitting, now what about the model itself ? The following sample shows how a model could be stored [33] in a CSVM table. The chosen model is very used in enzyme kinetics: a variation of Michaelis-Menten mechanism that is converted as an EDO system and stored with all needed parameters for numerical integration inside a CSVM table using a key/values scheme.

A model taken from enzymology :

$$E + S \underset{k_{-1}}{\overset{k_1}{\rightleftarrows}} ES \overset{k_2}{\longrightarrow} E + P$$

Converted in a EDO system :

$$\frac{d[S]}{dt} = k_{-1}\,[ES] - k_{+1}\,[E][S]$$

$$\frac{d[P]}{dt} = k_{+2}\,[ES]$$

…

The corresponding CSVM file shown in the following figure includes other equations (mass conservation, enzyme conservation), values for rate constants and molar concentrations, integration algorithm used (in this case RK4 stands for Runge-Kutta 4) and its parameters, other non-numerical data …

*Figure 22. – EDO system and parameters stored in a CSVM file (human readable form).*

```
ALGO    rk4     -                                -           -       EDO solver
TIME    0       1                                0.0000001   0.01    Seconds
SPEC    S       1.0e-3                           -           -       Substrat (M, publi= 10.10-6)
SPEC    P       0.0                              -           -       Product (M)
SPEC    ES      0.0                              -           -       Enzyme-substrat complex (M)
SPEC    E       1.0e-006                         -           -       Free Enzyme (M, publi= 45.5 10-12 M)
RATE    k1      2.0e+005                         -           -       ct. ES formation (M-1.s-1)
RATE    k-1     1.1e+003                         -           -       ct. ES disparition (s-1)
RATE    k2      900                              -           -       ct. P formation (s-1)
# Theoric KM of 10.10-3 M
PATH    S       <k -1>.ES - <k1>.E.S              -           -       dS/dt
PATH    P       <k2>.ES                          -           -       dP/dt
PATH    ES      <k1>.E.S - <k -1>.ES - <k2>.ES    -           -       dES/dt
PATH    E       - <k1>.E.S + <k -1>.ES + <k2>.ES  -           -       dE/dt
MONI    Cm      S+P+ES+E                         -           -       Matter conservation
MONI    Etotal  E+ES                             -           -       Enzyme conservation

#TITLE    Prob 3.3 of Shuler and Kargi S>>E
#HEADER   KEY     UNDEF   UNDEF   UNDEF   UNDEF   UNDEF
#TYPE     TEXT    TEXT    TEXT    TEXT    TEXT    TEXT
#WIDTH    50      50      50      50      50      50
```

This is a proposal for coding the model in order to ensure genericity (a same formalism is usable for all models) and to keep human readable data (a formalism closely related to EDO system). Perhaps it is not the best solution for presenting data closely from the syntax and calling arguments of the numerical function used to make numerical integration. But in two cases, the CSVM file will be easily parsed independently of model's complexity, and a one problem will be eliminated.

*Similar examples could be found in chemistry, environmental sciences, etc. In these cases, CSVM shows that it is possible to store in the* ***same format****, the data and intermediate data, the results, the models. This is useful in a lot of experimental sciences because the development effort for computer scientists is reduced and they can focus on the problem itself.*

[33] Stored and parsed. We have used this approach for simulation, not for modeling because we use mainly analytical solutions rather than numerical solutions in our work.

# 6. Limit the number of data formats you use!

We have spoken about the use of CSVM as a canonical format for RAW data conversion and intermediate results. Extending the concept to other kind of data is fairly immediate. In this section some samples taken from structural bioinformatics are given.

## 6.1. Map of molecular interactions

The previous section has shown that CSVM can be used to store results of calculations. This ensures that these results could be used by another program or could be aggregated to larger dataset without any modifications. This is also true for non-time dependant data, here is a case for a map of molar interaction between a ligand (COA standing for Coenzyme A) and the water (HOH) or amino acid residues (LYS, ARG, ALA, SER, …) of an enzyme (1KRU PDB structure), *courtesy from* Vincent Mariaule [34].
A software pipe generates a custom 2D map of interaction using different programs: LIGPLOT [35] package and related programs [36] [37], the results of different calculations are filtered, extracted, converted and aggregated in the final CSVM file.

*Figure 23. – A 2 D interaction map stored as a CSVM table.*

```
1KRU    1KRU    LYS     195     N       d       HOH     2053    O       a   2.83
1KRU    1KRU    COA     204     S1P     d       IPT     207     O6      a   3.14
1KRU    1KRU    HOH     2053    O       d       COA     204     O4A     a   2.92
1KRU    1KRU    COA     204     OAP     d       COA     204     N7A     a   2.52
1KRU    1KRU    ARG     183     NH2     d       COA     204     O5A     a   3.09
1KRU    1KRU    ARG     183     NE      d       COA     204     O5A     a   3.22
1KRU    1KRU    COA     204     N6A     d       ALA     160     O       a   2.95
1KRU    1KRU    ALA     160     N       d       COA     204     O9P     a   3.16
1KRU    1KRU    SER     142     N       d       COA     204     O5P     a   2.81
1KRU    1KRU    IPT     207     C6      -       COA     204     S1P     -   3.60
1KRU    1KRU    COA     204     C2A     -       VAL     177     CG1     -   3.84
1KRU    1KRU    COA     204     S1P     -       HIS     115     CD2     -   3.48
1KRU    1KRU    COA     204     C6A     -       ALA     175     CB      -   3.89
1KRU    1KRU    COA     204     C7P     -       TRP     139     CH2     -   3.76
1KRU    1KRU    COA     204     CEP     -       TRP     139     CH2     -   3.65
1KRU    1KRU    COA     204     CEP     -       TRP     139     CZ3     -   3.60
1KRU    1KRU    COA     204     C6P     -       TRP     139     CZ2     -   3.74
1KRU    1KRU    COA     204     C7P     -       TRP     139     CZ2     -   3.57
1KRU    1KRU    COA     204     C3P     -       ALA     105     CB      -   3.45

#TITLE    Ligplot summary!ligplot2csvm.py (v:1.04)!1KRU
#HEADER   PDB   CHAIN   RESNAME  RESSEQ    ATYPE   FROM    RESNAME   RESSEQ    ATYPE   TO      DIST
#TYPE     TEXT  TEXT    TEXT     INTEGER   TEXT    TEXT    TEXT      INTEGER   TEXT    TEXT    FLOAT
#WIDTH    10    10      10       10        10      10      10        10        10      10      10
```

The *d* and *a* acronyms stands for donor and acceptor (hydrogen interactions), the columns RESSEQ are used to store the residue number in the protein sequence or in the crystallographic structure. The columns ATYPE store the atom types implied in interactions. The formalism is closely related to the Protein Data Bank [38] (PDB) formalism. The last column is the distance between the two atoms found by row.
Is structural data (a PDB [39] file) or sequence data could be stored in CSVM files ? Definitely yes, but the interest could be limited because a lot of file formats exists and are widely used.

---

[34] V. Mariaule. Etude structurale des histones acétyl transférases. Rapport M1 BBT, Université Paul Sabatier Toulouse III (2008).

[35] A.C. Wallace, R.A. Laskowski, J.M. Thornton. LIGPLOT: a program to generate schematic diagrams of protein-ligand interactions. Protein Eng. (1995) 8:2, 127-134.

[36] HBPLUS: Hydrogen Bond Calculation Program - I.K. McDonald, J.M. Thornton. Satisfying Hydrogen Bonding Potential in Proteins (1994) JMB 238, 777-793.

[37] NACCESS: Solvent accessible area calculations – S. Hubbard, J. Thornton (1992).

[38] RCSB Protein Data Bank - http://www.rcsb.org/pdb/

[39] In the case of a PDB file, the coordinates PDB block is immediately transposable, in the case of CSVM the use of delimiters permits to add new columns in any order and to add useful annotations. For the header PDB block using CSVM could simplify the metadata organization because CSVM is not restricted to a particular number of columns or characters per row. Sometimes, for intermediate results in calculations or protein coordinate normalization we had used CSVM for coordinate block.

## 6.2. BLOSSUM and PAM matrixes

For people working in genomics or proteomics, BLOSUM or PAM substitution matrixes [40] are well known objects. The next figure shows a CSVM version of a BLOSSUM 62 matrix. For readability, some columns of the metadata block are missing (and marked by '…' characters at right).
The order of amino acid names (1-letter code in `#HEADER`) is the same in columns (left to right) and rows (top to bottom), data taken from Hiroto Saigo [41].

*Figure 24. – A Blossum 62 matrix encoded in a CSVM table.*

```
4     -     -     -     -     -     -     -     -     -     -     -     -     -     -     -     -     -     -     -
-1    5     -     -     -     -     -     -     -     -     -     -     -     -     -     -     -     -     -     -
-2    0     6     -     -     -     -     -     -     -     -     -     -     -     -     -     -     -     -     -
-2    -2    1     6     -     -     -     -     -     -     -     -     -     -     -     -     -     -     -     -
0     -3    -3    -3    9     -     -     -     -     -     -     -     -     -     -     -     -     -     -     -
-1    1     0     0     -3    5     -     -     -     -     -     -     -     -     -     -     -     -     -     -
-1    0     0     2     -4    2     5     -     -     -     -     -     -     -     -     -     -     -     -     -
0     -2    0     -1    -3    -2    -2    6     -     -     -     -     -     -     -     -     -     -     -     -
-2    0     1     -1    -3    0     0     -2    8     -     -     -     -     -     -     -     -     -     -     -
-1    -3    -3    -3    -1    -3    -3    -4    -3    4     -     -     -     -     -     -     -     -     -     -
-1    -2    -3    -4    -1    -2    -3    -4    -3    2     4     -     -     -     -     -     -     -     -     -
-1    2     0     -1    -3    1     1     -2    -1    -3    -2    5     -     -     -     -     -     -     -     -
-1    -1    -2    -3    -1    0     -2    -3    -2    1     2     -1    5     -     -     -     -     -     -     -
-2    -3    -3    -3    -2    -3    -3    -3    -1    0     0     -3    0     6     -     -     -     -     -     -
-1    -2    -2    -1    -3    -1    -1    -2    -2    -3    -3    -1    -2    -4    7     -     -     -     -     -
1     -1    1     0     -1    0     0     0     -1    -2    -2    0     -1    -2    -1    4     -     -     -     -
0     -1    0     -1    -1    -1    -1    -2    -2    -1    -1    -1    -1    -2    -1    1     5     -     -     -
-3    -3    -4    -4    -2    -2    -3    -2    -2    -3    -2    -3    -1    1     -4    -3    -2    11    -     -
-2    -2    -2    -3    -2    -1    -2    -3    2     -1    -1    -2    -1    3     -3    -2    -2    2     7     -
0     -3    -3    -3    -1    -2    -2    -3    -3    3     1     -2    1     -1    -2    -2    0     -3    -1    4

#TITLE   BLOSSUM62
#HEADER  A       R       N       D       C       Q       E       G       H       I       ...
#TYPE    NUMERIC NUMERIC NUMERIC NUMERIC NUMERIC NUMERIC NUMERIC NUMERIC NUMERIC NUMERIC ...
#WIDTH   10      10      10      10      10      10      10      10      10      10      ...
#META    from http://sunflower.kuicr.kyoto-u.ac.jp/~hiroto/project/optaa.html
```

*Using same and unified file format for different parameters sets is interesting to minimize error and developments. Standard or reference data/charts are used in a lot of scientific fields, and it could be useful to take advantage from a canonical and generic format.*

## 6.3. PDB hetero compound dictionary and variations

The following figure shows the bottom of the PDB's databank hetero dictionary [42] encoded in CSVM. This file of about 2000 rows, is used to store PDB and chemical component identifier correspondences.
In example, the chemical component encoded with name ZZM (1-propan-2-yl-3-(2-pyridin-3-ylethynyl)pyrazolo[4,5-e]pyrimidin-4-amine) is found in PDB structure 2WXM (phosphoinositide-3-OH kinase). So, the first column is for the chemical compound PDB code (3 chars) and the column 2 is for a list (values separated by a space) of PDB entries (4 chars) in which the compound is found.
The CSVM format is not locked to tabular data of a given number of columns, formerly the following table is constituted of two columns (HETNAME, INPDB) but in the second column, the values could be single (last lines shown) or multiple (first line with 3 values: **2wtv**, **2wtw**, **2x81**).
It is possible to make 3 columns to store these values, but if the average amount of values by line is near 1, you will have a lot of unused columns needed by lines in which the number of values is high. This is precisely the case in this example; and it is better to define lines with two columns and to use a secondary delimiter in the INPDB column. Here a ' ' (space) character is used, but another could be used, the best in this case is a character [43] not found in current scientific data.

40 Substitution matrix (Wikipedia) - http://en.wikipedia.org/wiki/Substitution_matrix
41 Akutsu Laboratory - http://sunflower.kuicr.kyoto-u.ac.jp/index.html.en
42 Ligand Expo Downloads - http://ligand-expo.rcsb.org/ld-download.html
43 The '§' could be used and the corresponding string with 3 values in INPDB column becomes '2wtv§2wtw§2x81'.

The CSVM parser split the file according to the main delimiter (generally a TAB) and the application program, read CSVM matrix cells and split them knowing the secondary (or ternary) delimiter, if multiple values are expectable inside.
The CSVM formalism allows to send a signal through the parser about that (multiple values in cells). The #META row could be used with a specific keyword inside (the signal and the secondary delimiter), in example: `'#META    MULTIPLE_VALUES_IN_CELLS[$] …'` or `'#META      CODE104 CODE108 …'` etc.
This signal can be interpreted (post-parsing) by the application layer. As usual we don't define in CSVM specification this kind of mechanism; the interest of CSVM is precisely to provide architecture for data and metadata but without any normalization.

*Figure 25. – PDB and chemical component identifier (column HETNAME) correspondences encoded in a CSVM file.*

```
...
ZZL 2wtv 2wtw 2x81
ZZM 2wxm
ZZN 2wxg
...
ZZU 2wbp 2wbq
ZZY 2wd1
ZZZ 2cfi

#TITLE  cc-to-pdb.tdd
#HEADER HETNAME INPDB
#TYPE   TEXT    TEXT
#WIDTH  10  10
#META   http://ligand-expo.rcsb.org/ld-download.html|20jul2010
```

In this sample the `#META` field is used to store the download location, and a release date (using another delimiter the `'|'` character). The CSVM format let the user free from defining characters used as primary or secondary delimiters, both for data or metadata block, the parser knows only column delimiter, others separators are used by application.

## Real case: how to merge CSVM tables

The previous file (Figure 25) is very close from original file (the dictionary released by the PDB): only a metadata block has been added. This is also the case for the next file (Figure 26): another dictionary taken from the PDB listing the same hetero compounds with the same 3 letter code (column HETNAME) and two other columns for a chemical formula written in SMILES format (SMI column) and a chemical name (MOLNAME):

*Figure 26. – Last rows of a SMILES data file including PDB chemical component identifier (column HETNAME).*

```
...
C(CNC(=N)N)C(C(C(=O)O)N)O                               ZZU   (2s,3s)-hydroxyarginine
c1ccc(c(c1)[N+](=O)[O-])S(=O)(=O)n2ccc3c2cc(cn3)C(=O)N   ZZY   1-[(2-nitrophenyl)sulfonyl]-1h-pyrrolo[3,2-
b]pyridine-6-carboxamide
C1C(NC2=C(N1)N=C(NC2=O)N)C=O                            ZZZ   6-formyltetrahydropterin

#TITLE  Components-smiles-oe.smi
#HEADER SMI HETNAME MOLNAME
#TYPE   TEXT    TEXT    TEXT
#WIDTH  10  10  10
#META   http://ligand-expo.rcsb.org/ld-download.html|20jul2010
```

Even if the only modifications were to add a metadata block, this enables complex operations such as the merging of the files from Figure 25 and of Figure 26. This can be done with a few lines of code and functions from the Python CSVM toolkit. This example is given in the following and it will be used to illustrate some subtleties related to the manipulation of RAW data sets.

After importation of some modules of the Python toolkit, the first step is to create a csvm_ptr Python class **c1** and to read the first CSVM file (Figure 25). The columns HETNAME and INPDB can exist (only one or more column than one) or not. The indices (column number beginning with zero) of 'HETNAME' and 'INPDB' are extracted from **c1** object, the number of rows of **c1** are also counted:

```
c1 = csvm_ptr()
c1 = csvm_ptr_read_extended_csvm(c1,"figure25.csvm","\t")
h1 = csvm_ptr_getcolind(c1,'HETNAME')
i1 = csvm_ptr_getcolind(c1,'INPDB')
r1 = c1.DATA_R
print "c1: column 'HETNAME' found %d times - first occurence at index [%d]" % (len(h1), h1[0])
print "c1: column 'INPDB' found %d times - first occurence at index [%d]" % (len(i1), i1[0])
print "c1: found %d rows" % (c1.DATA_R)
```

The same process is done for the second file (Figure 26). But before, a row INPDB is added in the second csvm_ptr class **c2**, at the right of the data block:

```
c2 = csvm_ptr()
c2 = csvm_ptr_read_extended_csvm(c2,"figure26.csvm","\t")
c2 = csvm_ptr_add_hash(c2,['INPDB'],['TEXT'],['10'],'')
h2 = csvm_ptr_getcolind(c2,'HETNAME')
i2 = csvm_ptr_getcolind(c2,'INPDB')
r2 = c2.DATA_R
print "c2: column 'HETNAME' found %d times - first occurence at index [%d]" % (len(h2), h2[0])
print "c2: column 'INPDB' found %d times - first occurence at index [%d]" % (len(i2), i2[0])
print "c2: found %d rows" % (r2)
```

The script output shows that the number of rows is not the same in **c1** (12107) and in (15071) **c2**:

```
c1: column 'HETNAME' found 1 times - first occurence at index [0]
c1: column 'INPDB' found 1 times - first occurence at index [1]
c1: found 12107 rows
c2: column 'HETNAME' found 1 times - first occurence at index [1]
c2: column 'INPDB' found 1 times - first occurence at index [3]
c2: found 15071 rows
```

This can be explained by the fact that a particular compound is known in PDB (Figure 26, **c2**) but not yet found as ligand or hetero compound inside a protein structure (Figure 25, **c1**).
In this case (**c1**.DATA_R < **c2**.DATA_R and **c2** is strictly a subset of **c1**) the choice is to extract values INPDB from **c1** and to fill column INPDB of **c2** when the same HETNAME value is found in the two data blocks.
For each value 'HETNAME' of **c2** a check is performed in **c1** (column HETNAME) in order to find if the same value exists. *If no*: the value is added to the **notfound** Python list, *if yes*: the counter **hetcount** is incremented. *If yes*: the corresponding row indice (**ind**) in **c1** column HETNAME is recorded.
The value of **ind** is used to get the INPDB value in the corresponding column of **c1**, and the value is then copied in **c2** at the current row indice (given by **i**) and in the corresponding column INPDB of **c2**:

```
"""
Extract the column 'HETNAME' from c1.
The column will be used to check if a 'HETNAME' value is found in c1
rather than using directly the data block.
"""

col = csvm_ptr_getcol(c1,'HETNAME')
hetcount = 0
notfound = []

"""
The loop: for each 'HETNAME' of c2
Check if 'HETNAME' value is found in column 'HETNAME' of c2.
If yes, get the indice of this value in c1 and add the 'INPDB' value
in the same row (of c1) to the current row and 'INPDB' column of c2
"""

for i in range (0, c2.DATA_R, 1):
   hetname2 = c2.DATA[i][h2[0]]
   if hetname2 in col[1]:
       ind = col[1].index(hetname2)
#       print "found %s at %d" % (hetname2, ind)
       c2.DATA[i][i2[0]] = c1.DATA[ind][i1[0]]
       hetcount += 1
   else:
#       print "not found %s" % (hetname2)
       notfound.append(hetname2)
```

```
print "c2: added %d hetnames" % (hetcount)
print "c1: %d hetnames not found" % (len(notfound))
```

The following script output shows that all HETNAME values of **c1** have been checked, and the summation of *added* and *not found* gives the good value of 15071 (number of rows for **c2** data block):

```
c2: added 12107 hetnames
c1: 2964 hetnames not found
```

Now it is possible to make an (optional) dump of the modified **c2**, to save a new CSVM file and to clean data structures in memory:

```
#c2.csvm_ptr_dump(0,0)
s = csvm_ptr_make_csvm(c2,"\n","\t")
file_str2file("result.csvm", s)
c1.csvm_ptr_clear()
c2.csvm_ptr_clear()
del(col[1])
```

The following table shows a part of the resulting CSVM file. The files taken for the programming example are more recent than the two used for previous figures. If a comparison is done with the file of Figure 25 we see obviously than 3 new compounds (ZZV, ZZW, ZZX) have been added:

*Table 1. – Result of table merging (only the 6 last rows are shown).*

| SMI | HETNAME | MOLNAME | INPDB |
|---|---|---|---|
| C(CNC(=N)N)C(C(C(=O)O)N)O | ZZU | (2s,3s)-hydroxyarginine | 2wbp 2wbq |
| CN(c1c2cccnc2c(c3c1CN(C3=O)Cc4ccc(cc4)F)O)S(=O)(=O)C | ZZV | n-[7-(4-fluorobenzyl)-9-hydroxy-8-oxo-7,8-dihydro-6h-pyrrolo[3,4-g]quinolin-5-yl]-n-methylmethanesulfonamide | 3oyd |
| c1ccc2c(c1)c3cc(ncc3n2Cc4ccc(cc4)F)C(=O)NO | ZZW | 9-(4-fluorobenzyl)-n-hydroxy-9h-beta-carboline-3-carboxamide | 3oyc |
| CCN1CC(n2c3c(c(c2C1=O)O)C(=O)N(N=C3C(=O)NC)Cc4ccc(c(c4)Cl)F)C | ZZX | (6s)-2-(3-chloro-4-fluorobenzyl)-8-ethyl-10-hydroxy-n,6-dimethyl-1,9-dioxo-1,2,6,7,8,9-hexahydropyrazino[1',2':1,5]pyrrolo[2,3-d]pyridazine-4-carboxamide | 3oyb 3oyj 3oyl 3oyn |
| c1ccc(c(c1)[N+](=O)[O-])S(=O)(=O)n2ccc3c2cc(cn3)C(=O)N | ZZY | 1-[(2-nitrophenyl)sulfonyl]-1h-pyrrolo[3,2-b]pyridine-6-carboxamide | 2wd1 |
| C1C(NC2=C(N1)N=C(NC2=O)N)C=O | ZZZ | 6-formyltetrahydropterin | 2cfi |

The next sample [44] is the output (using the csvm_ptr_dump method) of the first 5 rows of **c2** and it shows some compounds not found in **c1**. This is the case of compounds [000] and [005], the corresponding rows (indices 0 and 5) have the last field empty (no value between brackets).

*Figure 27. – Dump of the CSVM class corresponding to the Table.1 (only the 6 first rows of data block are shown).*

```
DUMP: CSVM info {
SOURCE result.csvm
CSV    CSVM
META   [http://ligand-expo.rcsb.org/ld-download.html]
TITLE_N     1
TITLE  Components-smiles-oe.smi with INPDB field added
HEADER_N    4
TYPE_N 4
WIDTH_N     4
0  10  TEXT    {SMI}
1  10  TEXT    {HETNAME}
2  10  TEXT    {MOLNAME}
3  10  TEXT    {INPDB}
DATA_R 15071
DATA_C 4
   15071   4

0  [COC(=O)O][000][methyl hydrogen carbonate][]
1  [COc1cc(cc(c1OC)OC)C(C(=O)N2CCCCC2C(=O)OC(CCCc3ccccc3)CCCc4cccnc4)(F)F][001][1-[2,2-difluoro-2-(3,4,5-trimethoxy-phenyl)-acetyl]-piperidine-2-carboxylic acid 4-phenyl-1-(3-pyridin-3-yl-propyl)-butyl ester][1j4r ]
2  [CCC(C)C(C(=O)NC(CC(C)C)C(=O)O)NC(=O)C(Cc1ccccc1)CC(=O)NO][002][n-[(2r)-2-benzyl-4-(hydroxyamino)-4-oxobutanoyl]-l-isoleucyl-l-leucine][2fv9 ]
```

[44] The rows of the data block (grey and gray background, black or yellow foreground) are wrapped. The row indices (yellow) at left and INPDB values (yellow) at right, compound names (HETNAME, yellow) in the middle of the row.

```
3   [CC(C)CN1c2c(c(n(n2)Cc3cccc4c3cccc4)c5ccncc5)C(=O)N(C1=O)C][003][5-methyl-7-(2-methylpropyl)-2-
(naphthalen-1-ylmethyl)-3-pyridin-4-yl-2h-pyrazolo[3,4-d]pyrimidine-4,6(5h,7h)-dione][2jfz ]
4   [c1ccc(cc1)C(C(=O)O)N][004][(2s)-amino(phenyl)ethanoic acid][1sm1 1yit 1yjw 2z2p ]
5   [c1ccc(cc1)CC(C(C(=O)O)O)N][005][(2s,3s)-3-amino-2-hydroxy-4-phenylbutanoic acid][]
```

This sample demonstrates that it is possible to make an operation that mixes the intersection and union (Section 4.3) of RAW tables in a simple manner.

Nevertheless this is a very favorable situation because all columns of **c1** have the same number of rows, no duplicates are found in HETNAME column of **c1** and **c2**, no blank values inserted between values … if it is not the case, the algorithm must be modified.

## Conclusion

This document is based on about ten years of CSVM usage in laboratory and does not cover all samples we have encountered. The examples of using CSVM are now numerous and all seem to demonstrate the plasticity of this format. This feature was essential to implement a SFMD (Single Format – Multiple Data) paradigm for tabular data.
This format was also found to operate very well with tabular like data (data that can be organized in a table, but are not inherently tabular) such as multiple key/values groups in the same data block.
The use of CSVM as a canonical format is of a great interest in a lot of processes involving manipulation of tables: files conversion, normalization of data and data types, intermediate files in software pipes, automatic processing of big tables with sparse or heterogeneous data.
CSVM is also a suitable format for gathering scientific objects and for exposition of data through a web framework, without any database system.

A generalized application could be the long term storage of RAW data. We have defined in this case the concept of 'data museum'. The data is stored with a minimal annotation but enough to ensure its reuse some time (month, years and more) after. This kind annotation is nearly the same as the tags on drawers (that contains minerals or fossils samples) of a collection. It provides a view on data, enough to enable a research of a sample by a specialist of the domain. It is not suitable in this case to embed data in a database system, because these data islands are too heterogeneous to be taken in account by a given system (that is not guaranteed to be functional for a long duration). In the same way, to focus efforts on RAW data is very important because, very often, all RAW data is not integrated exhaustively in databases. Later, these 'not used' data could be requested for future studies, but at this time these data will be lost. This risk is particularly encountered in environmental sciences (loss of 'zero time').
The use of CSVM in the short term could be also useful. Typically RAW data managers in laboratories cannot, in the same time, organize data collection; curate data and build a database. Perhaps in a lot of cases, a better approach could be to store CSVM files, and at demand to build (a RDBMS or another system) a specific view of the data space (particularly in order to respond to a scientific question). In this case, the CSVM files could give all information needed by IT teams to build the database, independently of the work of RAW data managers. A well curated flat database is often better (from a scientific point of view) than any RDBMS. The evolution of JavaScript frameworks and search engines shows also than a collection of CSVM files could be exposed using a web interface with non negligible search capabilities. So, CSVM could be useful to resolve the gap between people which wait a new database to store data, and people which wait data to model and build the same database (the use of AGILE techniques is not ever enough to solve this problem).
CSVM could also give support in the case of not published scientific data, because the CSVM format provides a way to store these tables (perhaps 80 per cent of data in a lot of experimental sciences) and allow a normalization effort, without too much complexity for the scientific community. Clearly, the research of funds in order to build databases for never (perhaps) used data could be a harder work than storing data in an adequate format.
Beyond of scientific use, this approach could be interesting in the Open Data. In this case, some questions are common with long term storage and unpublished data fields.

## Specific References

CSVM as a derivative of CSV is an Open file Format. The CSVM-1 specification and use of CSVM dictionaries are covered in the two following documents:

A CSVM toolkit (Python programming language, Open Source: CeCILL free software license agreement) is available from the corresponding author of this manuscript.

## Acknowledgments

I am grateful to Professor Pierre Tisnès (SPCMIB, UMR5068 CNRS/Université Paul Sabatier) and Dr Philippe Vervier (CESAC, GIS ECOBAG, Acceptables Avenirs, ECOLAB UMR 5245 CNRS/Université Paul Sabatier) for supporting the earlier phases of this work.
I am grateful to Dr Michel Baltas (SPCMIB, UMR 5068 CNRS/Université Paul Sabatier) for supporting this work.
*I would like to thank the CNRS, the "Université Paul Sabatier" and GIS ECOBAG for their financial support.*

I thank Dr Jean Louis Tichadou (Université Paul Sabatier); Professor Magali Gerino (ECOLAB, UMR 5245 CNRS/Université Paul Sabatier); Dr Philippe Vervier and members of their respective teams for advices and helpful discussions about this work.

I would like to thank all staff and students in different scientific fields and laboratories who contributed by providing data, writing code or manipulating RAW data, namely: Dr Pascal Hoffmann (staff, Chemistry) - Anne Laure Leomant (programmer) - Nathalie Gouardères (staff, Data Manager, Chemistry) - Mansi Trivedi (Bioinformatics) - Dr Daniel Redoulès (staff, Pierre Fabre Medicaments) - Dr Christian Lherbet (staff, Medicinal Chemistry) - Dr Cécile Baudoin-Dehoux (staff, Chemistry) - Dr Philippe Vervier (staff, Environmental Sciences) - Sarah Gimet (data manager, Environmental Sciences) - Gerome Beyries (programmer) - Jean-Olivier Butron (programmer) - Marjorie Catala (staff, data manager, Structural Biology) - Pr Magali Gerino (staff, Environmental Sciences) - Maya Lauriol (data manager, Environmental Sciences) - Dr Sabine Sauvage (staff, Environmental Sciences) - Dr Sabine Gavalda (Enzyme Kinetics) - Dr Casimir Blonski (staff, Molecular Enzymology) - Vincent Mariaule (data manager, Structural Biology).

## Annex-1

In this scheme the use of CSVM files makes easy the mapping of manual processing to automatic processing. The equations [45] [46] corresponding to these kinetic mechanisms (slow binding) are studied from the 60's and the enzymatic parameters can be resumed in the following way:

**Mechanism A**

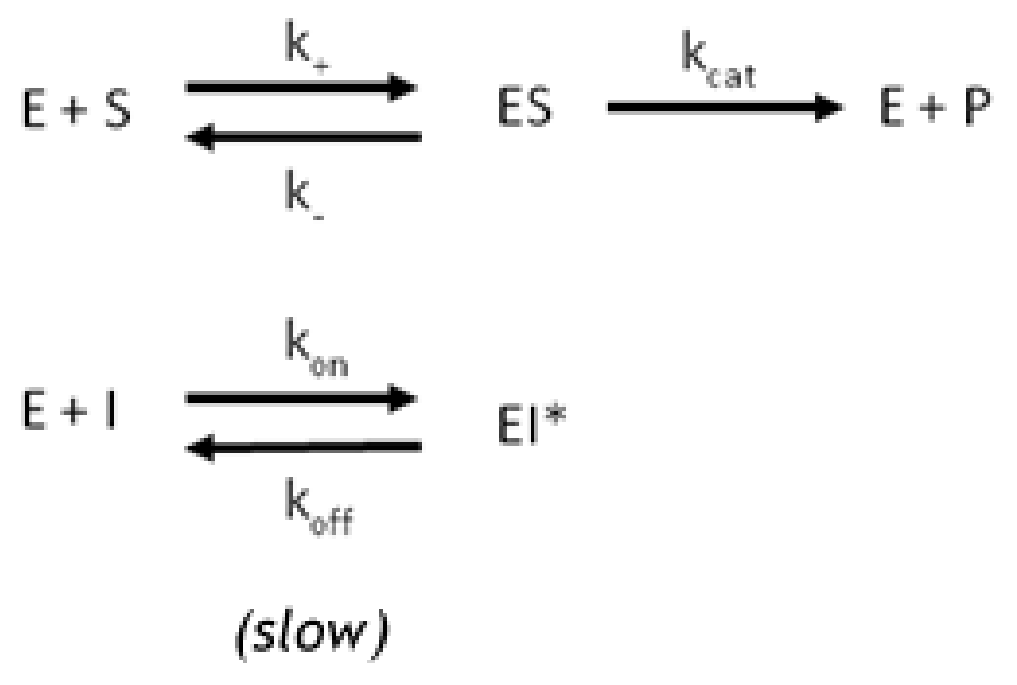

**Mechanism B**

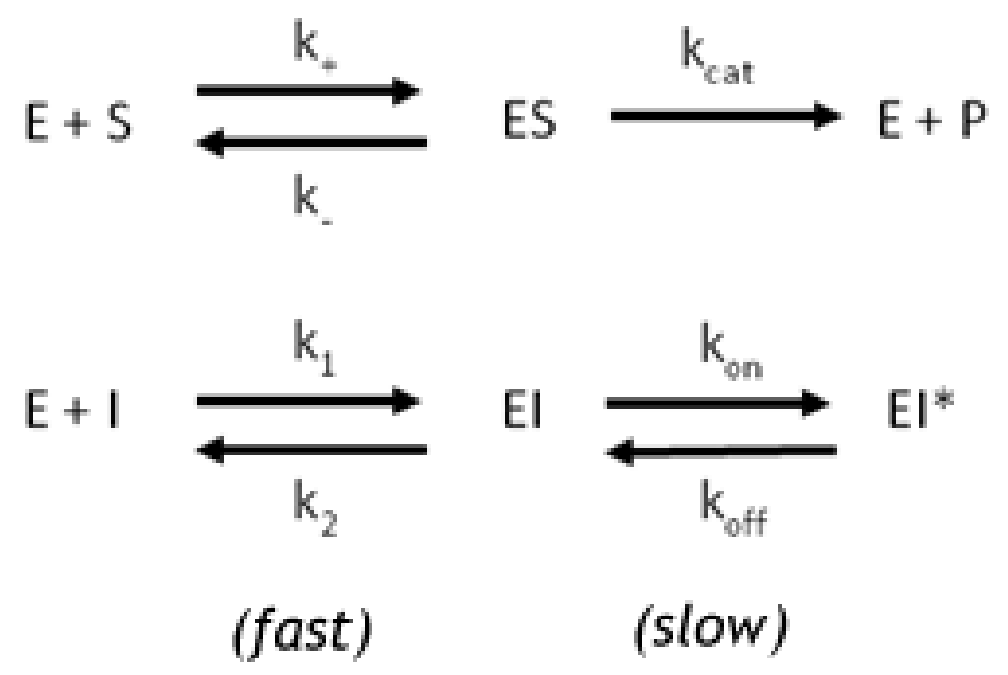

Mechanism A:

$$K_M = \frac{[E].[S]}{[ES]} = \frac{k_- + k_{cat}}{k_+}$$

$$K_I^* = \frac{[E].[I]}{[EI^*]} = \frac{k_{off}}{k_{on}}$$

Mechanism B:

$$K_M = \frac{[E].[S]}{[ES]} = \frac{k_- + k_{cat}}{k_+}$$

$$K_I^* = \frac{[E].[I]}{[EI]} = \frac{k_2}{k_1}$$

$$K_I^* = \frac{[E].[I]}{[EI] + [EI^*]} = \frac{K_I.k_{off}}{k_{on} + k_{off}}$$

For a given mechanism (i.e. the mechanism A) the mixing order of the compounds (Substrate, Enzyme, Inhibitor) in the reaction cell induces different progress curves (and corresponding equation schemes) for the variation of *[P]* vs. time *t*:

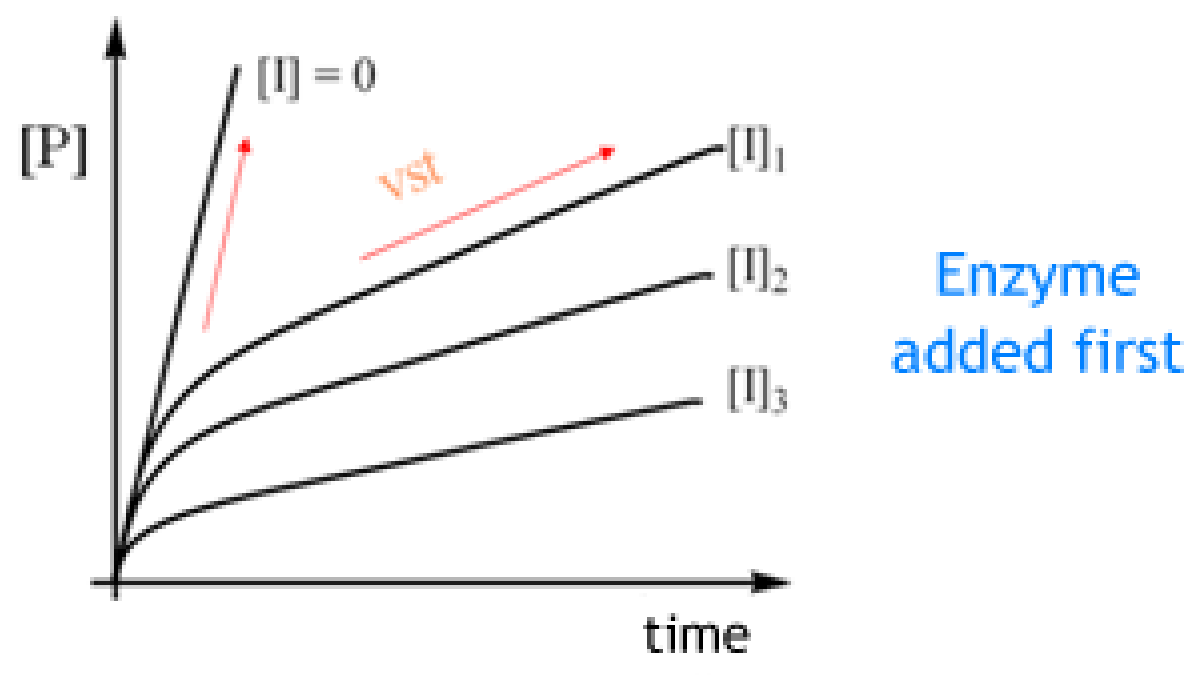

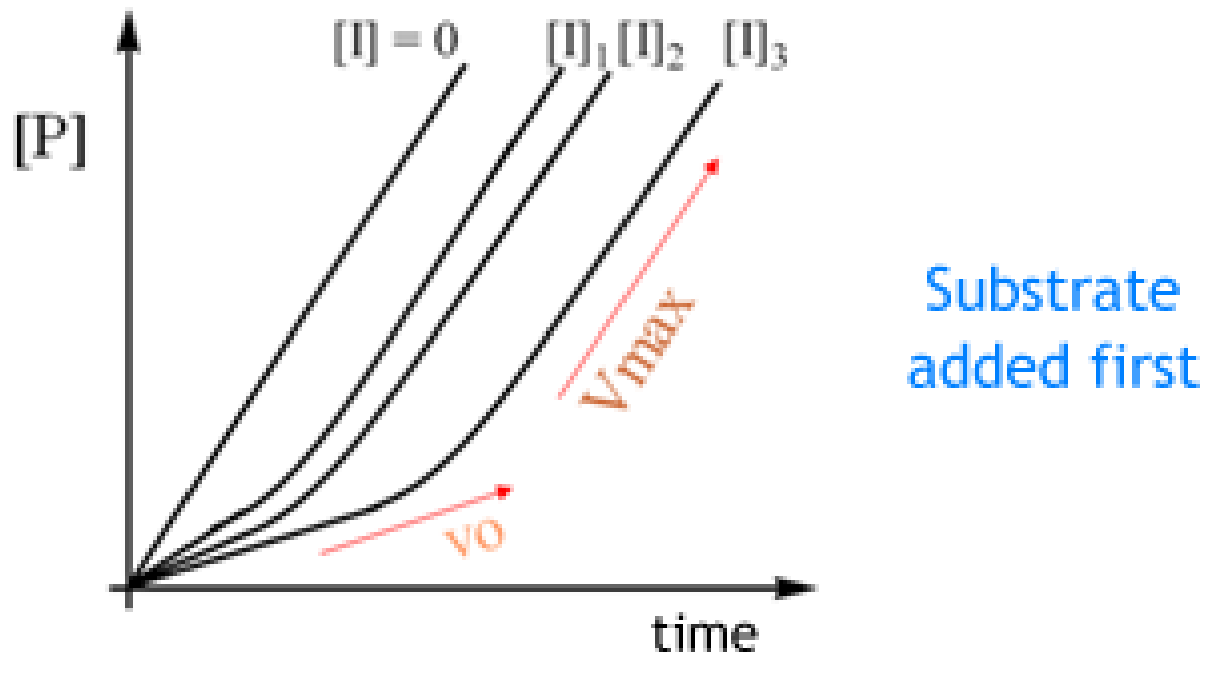

$$[P] = v_{st}.t + \frac{(V_0 - v_{st})(1 - e^{-k.t})}{k}$$

$$[P] = v_{max}.t + \frac{(V_0 - V_{max})(1 - e^{-k_{off}.t})}{k_{off}}$$

This is a particularity of slow binding mechanisms compared to more classic inhibition scheme found in enzyme kinetics science.

---

45 Morrison JF. Kinetics of the reversible inhibition of enzyme-catalysed reactions by tight-binding inhibitors. *Biochim Biophys Acta*. (1969) 185(2): 269–286.

46 Morrison JF, Walsh CT. The behavior and significance of slow-binding enzyme inhibitors. *Adv Enzymol Relat Areas Mol Biol.* (1988) 61: 201–301.

The analytic equations of $[P]=f(t)$ for each case of mixture show an apparent constant $k$ that can be computed by a direct fit of the progress curves (two column curves extracted from primary data). After calculation, this apparent constant found in secondary CSVM files and its value stored in key PSOL of the first key block (PNUM=1).
Knowing that the $K_M$ constant (characteristic of the Enzyme-Substrate affinity) is determined using another way (kinetics without inhibitor in the mixture) it is possible to compute the variation of $k_{on}$ , $k_{off}$ and $K_I$* using a linearization. Another curve fitting is done that uses the generated curves $k=f([I])$ stored in tertiary CSVM files and computed as a *secondary* calculation layer:

```
Secondary calculation layer sample
```

*1/v_st = f([I]) et k=f([I])*

$$[P] = v_{max}.t + \frac{(v_0 - V_{max})(1 - e^{-k_{off}.t})}{k_{off}}$$

$$k = k_{off} + k_{on} \frac{[I]}{1 + \frac{[S]}{K_M}}$$

```
Secondary calculation layer sample
```

*V_max/v_O = f([I]) ou v_O/(V_max – v_O) =f([I])*

$$\frac{V_{max}}{v_0} = 1 + \frac{[I]}{K_I^*}$$

$$\frac{v_0}{V_{max} - v_0} = \frac{K_I^*}{[I]}$$